\documentstyle[12pt,epsf]{article}

\renewcommand{\thefootnote}{\fnsymbol{footnote}}
\newlength{\pubnumber} 
\newcommand\pubblock[2]{\begin{flushright}\parbox{\pubnumber}
 {\begin{flushleft}#1\\ #2\\ \end{flushleft}}\end{flushright}}
\catcode`\@=11
\@addtoreset{equation}{section}
\def\theequation{\thesection.\arabic{equation}}
\def\section{\@startsection{section}{1}{\z@}{3.5ex plus 1ex minus .2ex}
 {2.3ex plus .2ex}{\large\bf}}
\def\subsection{\@startsection{subsection}{2}{\z@}{2.3ex plus .2ex}
 {2.3ex plus .2ex}{\bf}}
\newcommand\Appendix[1]{\def\thesection{Appendix \Alph{section}}
 \section{\label{#1}}\def\thesection{\Alph{section}}}

\def\MH#1#2#3{
\vev{H_{#1}} =
\left ( 
\begin{array}{c}
{#2} \\ {#3}
\end{array} 
\right ) }

\begin{document}

\begin{titlepage}
\samepage{
\setcounter{page}{1}
\rightline{ACT-04/01}
\rightline{BUHEP-01/01}
\rightline{CTP-TAMU-04/01}
\rightline{OUTP-01-08P}
\rightline{\tt hep-ph/0104091}
\rightline{April 2001}
\vfill
\begin{center}
 {\Large \bf Phenomenology of Non-Abelian Flat Directions\\ 
             in a Minimal Superstring Standard Model}
\vfill
\vskip .4truecm
\vfill {\large
        G.B. Cleaver,$^{1,2,3}$\footnote{gcleaver@physics.tamu.edu}
        A.E. Faraggi,$^{4}$\footnote{faraggi@thphys.ox.ac.uk}
        D.V. Nanopoulos$^{1,2,5}$\footnote{dimitri@soda.physics.tamu.edu},}

       {\large and
        J.W. Walker$^{1}$\footnote{jwalker@physics.tamu.edu}}
\\
\vspace{.12in}
{\it $^{1}$ Center for Theoretical Physics,
            Dept.\  of Physics, Texas A\&M University,\\
            College Station, TX 77843, USA\\}
\vspace{.06in}
{\it $^{2}$ Astro Particle Physics Group,
            Houston Advanced Research Center (HARC),\\
            The Mitchell Campus,
            Woodlands, TX 77381, USA\\}
\vspace{.06in}
{\it$^{3}$  Department of Physics, Baylor University,
            Waco, TX 76798-7316, USA\\}
\vspace{.06in}
{\it$^{4}$  Theoretical Physics, University of Oxford, 1 Keble Road,
            Oxford OX1 3NP, UK\\}
\vspace{.06in}
{\it$^{5}$  Academy of Athens, Chair of Theoretical Physics, 
            Division of Natural Sciences,\\
            28 Panepistimiou Avenue, Athens 10679, Greece\\}
\vspace{.025in}
\end{center}
\vfill
\begin{abstract}
Recently, we presented the first non-Abelian flat directions 
that produce from a heterotic string model solely 
the three-generation MSSM states as the massless spectrum in
the observable sector of the low energy effective
field theory.
In this paper we continue to develop the systematic techniques for 
the analysis of nonrenormalizable superpotential terms and non-Abelian
flat direction in realistic string models.
Some of our non-Abelian directions were $F$-flat
to {\it all} finite orders in the superpotential.
We study for the same string model the varying 
phenomenologies resulting from a large set
of such all-order flat directions. 
We focus on the quark, charged lepton, and Higgs doublet mass matrices
resulting for our phenomenologically superior non-Abelian flat direction.
We review and apply a string-related method for generating
large mass hierarchies
between MSSM generations, first discussed in string-derived 
flipped $SU(5)$ models,
when all generational mass terms are of 
renormalizable or very low non-renormalizable order.
\end{abstract}
\smallskip}
\end{titlepage}

\setcounter{footnote}{0}

\def\MH#1#2#3{
\vev{\H{#1}} =
\left( 
\begin{array}{c}
{#2} \\
{#3}
\end{array} \right) }
\def\at{ }
\def\beq{\begin{equation}}
\def\eeq{\end{equation}}
\def\beqn{\begin{eqnarray}}
\def\eeqn{\end{eqnarray}}
\def\no{\noindent }
\def\nolabel{\nonumber }

\def\NA{non-Abelian }

\def\gsim{{\buildrel >\over \sim}}
\def\lsim{{\buildrel <\over \sim}}

\def\ie{i.e., }
\def\eg{{\it e.g.}}
\def\eq#1{Eq.\ (\ref{#1}) }

\def\lt{<}

\def\slash#1{#1\hskip-6pt/\hskip6pt}
\def\slk{\slash{k}}

\def\dag{\dagger}
\def\qandq{\quad {\rm and} \quad} 
\def\qand{\quad {\rm and} } 
\def\andq{ {\rm and} \quad } 
\def\qwithq{\quad {\rm with} \quad} 
\def\qwith{ \quad {\rm with} } 
\def\withq{ {\rm with} \quad} 

\def\fhalf{\frac{1}{2}}
\def\fsqrt{\frac{1}{\sqrt{2}}}
\def\half{{\textstyle{1\over 2}}}
\def\third{{\textstyle {1\over3}}}
\def\quarter{{\textstyle {1\over4}}}
\def\sixth{{\textstyle {1\over6}}}
\def\phm{$\phantom{-}$}
\def\m{$\phantom{-}$}
\def\j{$-$}
\def\ps{{\tt +}}
\def\pps{\phantom{+}}

\def\zz{$Z_2\times Z_2$ }

\def\Tr{{\rm Tr}\, }
\def\tr{{\rm tr}\, }

\def\lam#1{\lambda_{#1}}
\def\non{\nonumber}
\def\smgg{ $SU(3)_C\times SU(2)_L\times U(1)_Y$ }
\def\smggb{ $SU(3)_C\times SU(2)_L\times U(1)_Y$}
\def\SM{Standard-Model }
\def\SUSY{supersymmetry }
\def\SSSM{supersymmetric standard model}
\def\MSSM{minimal supersymmetric standard model}
\def\MSSSM{MS$_{str}$SM }
\def\MSSSMc{MS$_{str}$SM, }
\def\obs{{\rm observable}}
\def\sig{{\rm singlets}}
\def\hid{{\rm hidden}}
\def\MS{M_{str}}
\def\Ms{$M_{str}$}
\def\MP{M_{P}}
\def\GeV{\,{\rm GeV}}
\def\TeV{\,{\rm TeV}}

\def\vev#1{\langle #1\rangle}
\def\mvev#1{|\langle #1\rangle|^2}
\def\mveV#1{|\langle #1\rangle|}

\def\UA{U(1)_{\rm A}}
\def\QA{Q^{(\rm A)}}
\def\mssm{SU(3)_C\times SU(2)_L\times U(1)_Y} 

\def\KM{Ka\v c-Moody }

\def\y{\,{\rm y}}
\def\l{\langle}
\def\r{\rangle}
\def\o#1{\frac{1}{#1}}

\def\zi{z_{\infty}}

\def\bc{\bar{c}}
\def\lh{\bar{h}}
\def\hb#1{\bar{h}_{#1}}
\def\bh#1{\bar{h}_{#1}}
\def\Htw{{\tilde H}}
\def\chibar{{\overline{\chi}}}
\def\qbar{{\overline{q}}}
\def\ibar{{\overline{\imath}}}
\def\jbar{{\overline{\jmath}}}
\def\Hbar{{\overline{H}}}
\def\Qbar{{\overline{Q}}}
\def\abar{{\overline{a}}}
\def\alphabar{{\overline{\alpha}}}
\def\betabar{{\overline{\beta}}}
\def\tautwo{{ \tau_2 }}
\def\thetatwo{{ \vartheta_2 }}
\def\thetathree{{ \vartheta_3 }}
\def\thetafour{{ \vartheta_4 }}
\def\ttwo{{\vartheta_2}}
\def\tthree{{\vartheta_3}}
\def\tfour{{\vartheta_4}}
\def\ti{{\vartheta_i}}
\def\tj{{\vartheta_j}}
\def\tk{{\vartheta_k}}
\def\calF{{\cal F}}
\def\smallmatrix#1#2#3#4{{ {{#1}~{#2}\choose{#3}~{#4}} }}
\def\ab{{\alpha\beta}}
\def\Minv{{ (M^{-1}_\ab)_{ij} }}
\def\ii{{(i)}}
\def\V{{\bf V}}
\def\N{{\bf N}}

\def\b{{\bf b}}
\def\S{{\bf S}}
\def\X{{\bf X}}
\def\I{{\bf I}}
\def\bone{{\mathbf 1}}
\def\bo{{\mathbf 0}}
\def\bs{{\mathbf S}}
\def\mS{{\mathbf S}}
\def\bS{{\mathbf S}}
\def\bb{{\mathbf b}}
\def\mb{{\mathbf b}}
\def\mX{{\mathbf X}}
\def\mI{{\mathbf I}}
\def\bI{{\mathbf I}}
\def\balpha{{\mathbf \alpha}}
\def\bbeta{{\mathbf \beta}}
\def\bgamma{{\mathbf \gamma}}
\def\bxi{{\mathbf \xi}}
\def\malpha{{\mathbf \alpha}}
\def\mbeta{{\mathbf \beta}}
\def\mgamma{{\mathbf \gamma}}
\def\mxi{{\mathbf \xi}}
\def\bphi{\overline{\Phi}}

\def\eps{\epsilon}

\def\t#1#2{{ \Theta\left\lbrack \matrix{ {#1}\cr {#2}\cr }\right\rbrack }}
\def\C#1#2{{ C\left\lbrack \matrix{ {#1}\cr {#2}\cr }\right\rbrack }}
\def\tp#1#2{{ \Theta'\left\lbrack \matrix{ {#1}\cr {#2}\cr }\right\rbrack }}
\def\tpp#1#2{{ \Theta''\left\lbrack \matrix{ {#1}\cr {#2}\cr }\right\rbrack }}
\def\l{\langle}
\def\r{\rangle}

\def\op#1{$\Phi_{#1}$}
\def\opp#1{$\Phi^{'}_{#1}$}
\def\opb#1{$\overline{\Phi}_{#1}$}
\def\opbp#1{$\overline{\Phi}^{'}_{#1}$}
\def\oppb#1{$\overline{\Phi}^{'}_{#1}$}
\def\oppx#1{$\Phi^{(')}_{#1}$}
\def\opbpx#1{$\overline{\Phi}^{(')}_{#1}$}

\def\oh#1{$h_{#1}$}
\def\ohb#1{${\bar{h}}_{#1}$}
\def\ohp#1{$h^{'}_{#1}$}

\def\oQ#1{$Q_{#1}$}
\def\odc#1{$d^{c}_{#1}$}
\def\ouc#1{$u^{c}_{#1}$}

\def\oL#1{$L_{#1}$}
\def\oec#1{$e^{c}_{#1}$}
\def\oNc#1{$N^{c}_{#1}$}

\def\oH#1{$H_{#1}$}
\def\oV#1{$V_{#1}$}
\def\oHs#1{$H^{s}_{#1}$}
\def\oVs#1{$V^{s}_{#1}$}

\def\p4{\Phi_4}
\def\pp4{\Phi^{'}_4}
\def\pb4{\bar{\Phi}_4}
\def\ppb4{\bar{\Phi}^{'}_4}
\def\p#1{{\Phi_{#1}}}
\def\P#1{{\Phi_{#1}}}
\def\pp#1{{\Phi^{'}_{#1}}}
\def\pb#1{{{\overline{\Phi}}_{#1}}}
\def\bp#1{{{\overline{\Phi}}_{#1}}}
\def\pbp#1{{{\overline{\Phi}}^{'}_{#1}}}
\def\ppb#1{{{\overline{\Phi}}^{'}_{#1}}}
\def\bpp#1{{{\overline{\Phi}}^{'}_{#1}}}
\def\bi#1{{{\overline{\Phi}}^{'}_{#1}}}
\def\ppx#1{{\Phi^{(')}_{#1}}}
\def\pbpx#1{{\overline{\Phi}^{(')}_{#1}}}

\def\h#1{h_{#1}}
\def\hb#1{{\bar{h}}_{#1}}
\def\hp#1{h^{'}_{#1}}

\def\Q#1{Q_{#1}}
\def\dc#1{d^{c}_{#1}}
\def\uc#1{u^{c}_{#1}}

\def\L#1{L_{#1}}
\def\ec#1{e^{c}_{#1}}
\def\Nc#1{N^{c}_{#1}}

\def\H#1{H_{#1}}
\def\V#1{V_{#1}}
\def\Hs#1{{H^{s}_{#1}}}
\def\sH#1{{H^{s}_{#1}}}
\def\Vs#1{{V^{s}_{#1}}}
\def\sV#1{{V^{s}_{#1}}}

\def\fdtv{FD2V }
\def\fdtp{FD2$^{'}$ }
\def\fdtpv{FD2$^{'}$v }

\def\FD2pv{FD2$^{'}$V }
\def\FD2p{FD2$^{'}$ }


\def\inbar{\,\vrule height1.5ex width.4pt depth0pt}

\def\IC{\relax\hbox{$\inbar\kern-.3em{\rm C}$}}
\def\IQ{\relax\hbox{$\inbar\kern-.3em{\rm Q}$}}
\def\IR{\relax{\rm I\kern-.18em R}}
 \font\cmss=cmss10 \font\cmsss=cmss10 at 7pt
 \font\cmsst=cmss10 at 9pt
 \font\cmssn=cmss9

\def\IZ{\relax\ifmmode\mathchoice
 {\hbox{\cmss Z\kern-.4em Z}}{\hbox{\cmss Z\kern-.4em Z}}
 {\lower.9pt\hbox{\cmsss Z\kern-.4em Z}}
 {\lower1.2pt\hbox{\cmsss Z\kern-.4em Z}}\else{\cmss Z\kern-.4em Z}\fi}

\def\Io{\relax\ifmmode\mathchoice
 {\hbox{\cmss 1\kern-.4em 1}}{\hbox{\cmss 1\kern-.4em 1}}
 {\lower.9pt\hbox{\cmsss 1\kern-.4em 1}}
 {\lower1.2pt\hbox{\cmsss 1\kern-.4em 1}}\else{\cmss 1\kern-.4em 1}\fi}

\hyphenation{su-per-sym-met-ric non-su-per-sym-met-ric}
\hyphenation{space-time-super-sym-met-ric}
\hyphenation{mod-u-lar mod-u-lar-in-var-i-ant}


\section{Minimal Superstring Standard Models}

Supersymmetry and heterotic string unification continue to be the
only extensions of the Standard Model\footnote{Standard Model here also
refers to its extensions that include neutrino masses.} that are 
motivated by a successful experimental prediction. Indeed, for over
a decade now, the consistency of the MSSM gauge coupling 
unification with the low energy experimental data \cite{mssmgcu} has
provided the most appealing hint for the validity of the big desert
scenario and Planck scale unification \cite{ross}.
On the other hand, heterotic string theory provides
a consistent theoretical framework to study how the 
structure of the Standard Model may arise from Planck scale physics.
The success of the MSSM gauge coupling unification implies
that the string models that should be constructed are those
that contain solely the MSSM spectrum in the effective 
low energy field theory below the string scale. 
The complete solution will then be achieved by embedding such models
in M-theory and extrapolating to the strong coupling regime
in which the string and MSSM unification scales coincide \cite{witten}.
It is therefore remarkable that recently we were able to construct
string models in which the only Standard Model charged fields, 
in the low energy effective field theory, coincide exactly with
the matter content of the MSSM \cite{cfn1,cfn2,cfn3,cfn4}. 
The starting point for the construction of the first known example of 
a ``Minimal Standard Heterotic-String Model,'' (MSHSM)
is the free fermionic 
string model first presented in ref.\ \cite{fny}, referred to as 
the ``FNY'' model. The FNY observable
gauge group is $SU(3)_C \times SU(2)_L \times U(1)_Y \times \prod_i U(1)_i$.
As with all three-generations $(2,0)$ string-based $SU(3)_C \times SU(2)_L 
\times U(1)_Y$ models possessing free-fermions, free lattice
bosons, or orbifolds, for internal degrees of freedom, FNY possesses an 
anomalous $U(1)_A$.
Cancellation of this anomaly via the standard string
mechanism and related appearance of the Fayet-Iliopoulos (FI) $D$-term
results in the breaking of supersymmetry near the string scale, 
unless an appropriate set of fields carrying the anomalous charge
take on vacuum expectation values (VEVs) that together cancel the effects of
the FI term, while keeping the respective VEVs of all other non-anomalous 
$D$-terms at zero. $F$-flatness must also be retained up to an order
in the superpotential that is consistent with observable sector
supersymmetry being maintained down to near the electroweak (EW) scale. 
Depending on the string coupling strength, $F$-flatness cannot be broken
by terms below eighteenth to twentieth
order\footnote{As coupling strength increases,
so does the required order of flatness.}. 

The eventual determination of the set of fields  
chosen to acquire VEVs 
(a.k.a.\  the flat direction) in a given string model, will be fixed by
non-perturbative effects. However, the set of perturbatively allowed flat directions
can be determined. The varying phenomenology of these flat directions can
be studied, or conversely, the subset of flat directions satisfying known 
phenomenological requirements can be focused on.
Initially we investigated FNY flat directions composed solely of VEVs
of non-Abelian singlet fields \cite{cfn1,cfn2,cfn3}.
While these directions offered a good first step in constructing
a physically viable MSHSM model, we found some 
short-comings to this approach. Rather, the inclusion of hidden-sector
non-Abelian fields was strongly suggested.
The necessity to include non-Abelian flat directions
was also suggested in analysis of other semi-realistic
free fermionic string models \cite{namix}.
These were, however, preliminary studies and 
the inclusion of the non-Abelian fields in the systematic analysis
of the flat directions is still lacking. 
Thus, in \cite{cfn4} we
began our study of FNY non-Abelian flat directions, which we continue 
in this paper. We emphasize that the main aim of the present
paper is to continue to develop the techniques and methodology that are
needed to confront string theory with the low energy experimental
data. We therefore stress that we do not intend to suggest or imply that
we will find the ``theory of everything'' through investigation of the
FNY model 
or of its MSHSM-producing flat directions. Nevertheless, the case may still 
well be that some of the general properties and phenomenology of the true
string vacuum will be gleaned in the process. 

Our paper is organized as follows. 
First, in Section 2 we review the general structure of free fermionic heterotic
string models.
Then in Section 3, we summarize the generic properties of Abelian and
non-Abelian flat directions and their application to heterotic string models.
We also present and discuss some of the results of our recent
non-Abelian flat direction search, including the constraints
we placed on this search.
Next, in Section 4 we delve into  \NA flat directions 
for the FNY model. 
In Section 5 we investigate the phenomenology associated with these directions.
Specifically we investigate the three generation
quark and lepton mass matrices, the Higgs effective $\mu$-terms,
and issue of proton decay in the FNY MSHSMs. 
One especially interesting feature we find is that, because of 
the several, initially massless, Higgs doublets in FNY, 
a large hierarchy between the 
three generations MSSM fields is possible even when the quark
and lepton mass terms are all of very low order in the
superpotential. 
While only one Higgs doublet pair eigenstate can survive in the low 
energy MSHSM effective field theory, that eigenstate
may have several non-eigenstate Higgs doublets 
as unequally weighted components. Because of the unequal weighting,
differing naturally by several orders of magnitude, and preferred
coupling between generations and Higgs components, a large
hierarchy can arise. 
In Section 6 we present our concluding remarks.  

\section{General structure of free fermionic models} 

In this section we briefly review the general features,
which reveal why non-Abelian flat directions
are in fact necessary for obtaining viable fermion
mass spectrum in the realistic free fermionic models.
The details of the spectrum and more elaborate discussions
on the general structure of the models, and the general methodology of
their analysis, are given in the references and are not repeated here. 

The free fermionic models are constructed by specifying a set of
boundary condition basis vectors for the world-sheet free fermions
\cite{fff}. The basis vectors that generate the realistic models can
be divided into two groups. The first consist of the NAHE
set\footnote{This set was first utilized by Nanopoulos, Antoniadis,
Hagelin and Ellis (NAHE) in the construction of the
flipped $SU(5)$\cite{fsu5}. NAHE={\it pretty} in Hebrew.} \cite{nahe},
$\{{\bf 1}, \mS, \mb_1,\mb_2,\mb_3\}$, 
and correspond to $\IZ_2\times \IZ_2$ orbifold 
compactification, where the three sectors $\mb_1$, $\mb_2$ and $\mb_3$
correspond to the three twisted sectors of the orbifold models.
The gauge group at this level is $SO(10)\times SO(6)^3\times E_8$,
and the models contain 48 multiplets in the $\mathbf 16$ representation  of 
$SO(10)$. The NAHE set divides the internal world-sheet fermions 
into several groups with the set of 12 internal world-sheet fermions,
$\{y,w\vert{\bar y},{\bar\omega}\}$, 
corresponding to the six dimensional ``compactified space'',
the 16 complex fermionic states 
$\{{\bar\psi}^{1,\cdots,5},{\bar\eta}^{1,2,3},{\bar\phi}^{1,\cdots,8}\}$
corresponding to the gauge sector, and $\chi^{1,\cdots,6}$
corresponding to the RNS fermions, of the orbifold model.
The $\{y,w\vert{\bar y},{\bar\omega}\}$ world-sheet fermions
are further divided into three cyclically symmetric groups,
which reflects the structure of the $\IZ_2\times \IZ_2$ orbifold.
The states from each of the twisted sectors, $\mb_1$, $\mb_2$ or $\mb_3$
are then charged with respect to one of these orbifold planes.
The NAHE set is common to a large class of realistic free
fermionic models. Among them are the flipped $SU(5)$ string models \cite{fsu5},
the Pati-Salam string models \cite{alr}, the String standard-like models
\cite{fny,slm}, and the left-right symmetric string models \cite{cfs}.
As we elaborate further below, this general $\IZ_2\times \IZ_2$
orbifold structure is the reason for the necessity to incorporate non-Abelian
flat directions. 

The second step in the basis construction consists of adding
to the NAHE set three, or four, additional boundary condition basis vectors,
typically denoted as $\{\alpha,\beta,\gamma\}$.
The effect of these is to break the $SO(10)$ symmetry to one of its
subgroups, and at the same time reduce the number of generations  
to three, one from each twisted sector $\mb_i$ $(i=1,2,3)$. There are several
important features of this class of models. The first is that since
the generations are obtained from the three distinct twisted
sectors of the $\IZ_2\times \IZ_2$ orbifold, each generation is charged
with respect to a distinct set of horizontal charges, which also holds
for the untwisted Higgs doublets. Thus, to generate fermion mass
terms that mix between the generations we need fields that are 
charged simultaneously with respect to at least two orbifold planes. 
Furthermore, because the charges of the generations under
the horizontal symmetries are $\pm1/2$ also, the charges of the 
mixing fields should combine to $\pm1/2$.

To understand how the fields with the required properties arise in the 
free fermionic models, one has to examine the hidden sector of
the free fermionic string models. The hidden sector is obtained
from the $E_8$ factor at the level of the NAHE set and is generated
by the $\{{\bar\phi}^{1,\cdots,8}\}$ right-moving world-sheet fermions.
The addition of the vector $2\gamma$ to the NAHE set induces the
symmetry breaking $E_8\rightarrow SO(16)$. At the same time the
sectors $\mb_j+2\gamma$ produces multiplets in the vectorial $\mathbf 16$ 
representation
of $SO(16)$. The basis vectors beyond the NAHE set further break the
$SO(16)$ symmetry to one of its subgroups, and the hidden sector
massless states from the sectors $\mb_j+2\gamma$ transform in non-Abelian
representations of the unbroken subgroup. 
It is precisely these states that induce the mixing terms between
the chiral states from the twisted sectors $\mb_i$ $(i=1,2,3)$.
These mixing terms therefore have the generic form 
${\mathbf 16}_i {\mathbf 16}_j\, {\mathbf 10}\, {\mathbf 16}_i
{\mathbf 16}_j\phi^n$,
where the first two $\mathbf 16$ are in the spinorial representation of the observable 
$SO(10)$, the $\mathbf 10$ is in the vector representation of the observable 
$SO(10)$ and correspond to the untwisted Higgs fields, the last
two $\mathbf 16$ are in the vector representation of the hidden $SO(16)$,
and $\phi^n$ is a combination of $SO(10)\times SO(16)$ scalar singlets.
We therefore see that, in general, producing fermion mass matrices 
with nontrivial structures necessitates that the hidden sector
non-Abelian states from the sectors $\mb_j+2\gamma$ obtain non-trivial
VEVs.

\section{Non-Abelian Flat Directions and Spacetime \\Supersymmetry} 

\subsection{$D$- and $F$-Flatness Constraints}

In \cite{cfn1,cfn2,cfn3,cfn4} we reviewed $D$- and 
$F$-flatness and their well known requirements for maintenance of 
spacetime supersymmetry. Thus, we will only summarize this here, but with a new emphasis on geometric interpretation of the $SU(2)$ non-Abelian VEVs.

Spacetime supersymmetry is broken in a model
when the expectation value of the scalar potential,
\beqn
 V(\varphi) = \half \sum_{\alpha} g_{\alpha} 
(\sum_{a=1}^{{\rm {dim}}\, ({\cal G}_{\alpha})} D_a^{\alpha} D_a^{\alpha}) +
                    \sum_i | F_{\varphi_i} |^2\,\, ,
\label{vdef}
\eeqn
becomes non-zero. 
The $D$-term contributions in (\ref{vdef}) have the form,   
\beqn
D_a^{\alpha}&\equiv& \sum_m \varphi_{m}^{\dagger} T^{\alpha}_a \varphi_m\,\, , 
\label{dtgen} 
\eeqn
with $T^{\alpha}_a$ a matrix generator of the gauge group ${\cal G}_{\alpha}$ 
for the representation $\varphi_m$.  
The $F$-term contributions are, 
\beqn
F_{\Phi_{m}} &\equiv& \frac{\partial W}{\partial \Phi_{m}} \label{ftgen}\,\, . 
\eeqn
The $\varphi_m$ are (spacetime) scalar superpartners     
of the chiral spin-$\half$ fermions $\psi_m$, which together  
form a superfield $\Phi_{m}$.
Since all of the $D$ and $F$ contributions to (\ref{vdef}) 
are positive semidefinite, each must have 
a zero expectation value for supersymmetry to remain unbroken.

For an Abelian gauge group, the $D$-term (\ref{dtgen}) simplifies to
\beqn
D^{i}&\equiv& \sum_m  Q^{(i)}_m | \varphi_m |^2 \label{dtab}\,\,  
\eeqn
where $Q^{(i)}_m$ is the $U(1)_i$ charge of $\varphi_m$.  
When an Abelian symmetry is anomalous, that is,
the trace of its charge 
over the massless fields is non-zero, 
\beqn
\Tr Q^{(A)}\ne 0\,\, ,
\label{qtnz}
\eeqn 
the associated $D$-term acquires a Fayet-Iliopoulos (FI) term,
$\eps\equiv\frac{g^2_s M_P^2}{192\pi^2}\Tr Q^{(A)}$, 
\beqn
D^{(A)}&\equiv& \sum_m  Q^{(A)}_m | \varphi_m |^2 
+ \eps \, .
\label{dtaban}  
\eeqn  
$g_{s}$ is the string coupling and $M_P$ is the reduced Planck mass, 
$M_P\equiv M_{Planck}/\sqrt{8 \pi}\approx 2.4\times 10^{18}$ GeV.
It is always possible to place the total anomaly into a single $U(1)$. 

The FI term breaks supersymmetry near the string scale,
\beqn 
V \sim g_{s}^{2} \eps^2\,\, ,\label{veps}
\eeqn  
unless its can be cancelled by a set of scalar VEVs, $\{\vev{\varphi_{m'}}\}$, 
carrying anomalous charges $Q^{(A)}_{m'}$,
\beq
\vev{D^{(A)}}= \sum_{m'} Q^{(A)}_{m'} |\vev{\varphi_{m'}}|^2 
+ \eps  = 0\,\, .
\label{daf}
\eeq
To maintain supersymmetry, a set of anomaly-cancelling VEVs must 
simultaneously be $D$-flat 
for all additional Abelian and the non-Abelian gauge groups, 
\beq
\vev{D^{i,\alpha}}= 0\,\, . 
\label{dana}
\eeq
The consistent solution of (\ref{dtab}), when placed into 
(\ref{daf}), specifies the overall VEV ``FI-scale'', $\vev{\alpha}$, of
the model.  A typical FNY value is $\vev{\alpha} \approx 7 \times 10^{16}$ GeV.

For the case of $SU(2)$, $T^{SU(2)}_a$ will take on the values of the three Pauli matrices,
\beq
\sigma_x =
\left (                    
\begin{array}{cc}
0 & 1 \\
1 & 0 \\
\end{array} \right ), \,\,
\sigma_y =
\left (
\begin{array}{cc}
0 & -i \\
i &  0 \\
\end{array} \right ), \,\,
\sigma_z =
\left (
\begin{array}{cc}
1 & 0 \\
0 & -1 \\
\end{array} \right ).
\label{pauli}
\eeq
Each component of the vector $\vec{D}$ in this internal space will be the total,
 summed over all fields of the gauge group, ``spin expectation value'' in the given direction.  Vanishing of the $\vev{\vec{D}\cdot\vec{D}}$ contribution to
$\vev{V}$ demands that $SU(2)$ VEVs be chosen such that the total
$\hat{x}, \hat{y},$ and $\hat{z}$ expectation values are zero.
The normalization length, $S^{\dagger}S$, of a ``spinor'' $S$ will generally be restricted to integer units by Abelian $D$-flatness constraints from the Cartan
 sub-algebra and any extra $U(1)$ charges carried by the doublet 
(cf.\  Eq.\  \ref{dtab} with $S^{\dagger}S$ 
playing the role of $| \varphi |^2$).
Each spinor then has a length and direction associated with it 
and $D$-flatness requires the sum, placed tip-to-tail, to be zero.
This reflects the generic non-Abelian $D$-flatness 
requirement that the norms of non-Abelian field VEVs are in a one-to-one 
association with a ratio of powers of a corresponding \NA gauge 
invariant \cite{luty95}.   

It will be useful to have an explicit (normalized to $1$) representation for 
$S(\theta, \phi)$.  This may be readily obtained by use of the rotation matrix,
\beq
R(\vec{\theta}) \equiv
e^{-i\frac{\vec{\theta}\cdot\vec{\sigma}}{2}} = 
\cos(\frac{\theta}{2}) -i \hat{\theta}\cdot\vec{\sigma}\sin(\frac{\theta}{2})\,\, ,
\label{rotmax}
\eeq
to turn 
$\left (
\begin{array}{c}
1 \\
0 \\
\end{array} \right )$
$\equiv \vert +\hat{z}\rangle$ through an angle $\theta$ about the axis
$\hat{\theta} = - \hat{i} \sin(\phi) + \hat{j} \cos{\phi}$.  The result,
$\left (
\begin{array}{c}
\cos{\frac{\theta}{2}} \\
\sin{\frac{\theta}{2}} \, e^{i \phi} \\
\end{array} \right )$
, is only determined
up to a phase and the choice
\beq
S(\theta, \phi) \equiv
\left ( 
\begin{array}{c}
\cos{\frac{\theta}{2}} \, e^{-i \frac{\phi}{2}} \\
\sin{\frac{\theta}{2}} \, e^{+i \frac{\phi}{2}} \\
\end{array} \right )
\label{spinor}
\eeq
will prove more convenient in what follows.  Within the range of physical 
angles, $\theta = 0 \rightarrow \pi$ and $\phi = 0 \rightarrow 2\pi$, each 
spinor configuration is unique (excepting $\phi$ phase 
freedom for $\theta = 0,\pi$) and
carries a one-to-one geometrical correspondence.  
Up to a complex coefficient,
the most general possible doublet is represented.

A non-trivial superpotential $W$ additionally imposes numerous constraints on allowed
sets of anomaly-cancelling VEVs, through the $F$-terms in (\ref{vdef}).
$F$-flatness (and thereby supersymmetry) can be broken through an 
$n^{\rm th}$-order $W$ term containing $\Phi_{m}$ when all of the additional 
fields in the term acquire VEVs,
\beqn
\vev{F_{\Phi_m}}&\sim& \vev{{\frac{\partial W}{\partial \Phi_{m}}}} 
      \sim \lambda_n \vev{\varphi}^2 (\frac{\vev{\varphi}}{\MS})^{n-3}\,\, ,
\label{fwnb2}
\eeqn
where $\varphi$ denotes a generic scalar VEV.
If $\Phi_{m}$ also carries a VEV, then
supersymmetry can be broken simply by $\vev{W} \ne 0$\footnote{The
lower the order of an $F$-breaking term, 
the closer the supersymmetry breaking scale 
is to the string scale.}.

\subsection{World Sheet Selection Rules\label{rizrules}}

To survive in a generic field theory, a term contributing to the potential
need only satisfy the sum of all charges equal to zero (thus ensuring
conservation of each Abelian charge) and appropriate gauge invariant pairings
of non-Abelian fields.  String theory, on the other hand, is much more
restrictive, demanding that each collection of fields pass a
``picture-changing'' test in order to conserve global worldsheet and
Ising charges and retain modular invariance.
This extra constraint is beneficial to our program by removing,
at an early stage, many $W$- and $F$-terms which would otherwise be dangerous
to supersymmetry.

In free fermionic heterotic string theory, extra world-sheet fermions are used
to cancel the four-dimensional conformal anomaly.  Fermionic boundary
conditions define a given model and physical states may be produced out
of either the Ramond (twisted) or Neveu-Schwarz (untwisted) sectors. In $N = 1$
supersymmetric models, six of the real left-moving fermions are bosonized,
creating three $U(1)$ categories into which all physical states are separated.
General selection rules predicting nonvanishing correlators have been formulated
in \cite{nansr} and in \nolinebreak[4] \cite{rizos91}.  For completeness and because
they are referenced in Section \ref{rizos}, these rules are here
(with alterations in form) restated.  The letter $n$ will designate the total
number of fields in a term.

\begin{enumerate}
\item Ramond fields must be distributed equally, mod $2$, among all
categories, {\it and}
\item For $n=3$, either\, :
\begin{enumerate}
\item There is $1$ field from each $R$ category.
\item There is $1$ field from each $NS$ category.
\item There are $2 R$ and $1 NS$ in a single category.
\end{enumerate}
\item For $n>3$\, :
\begin{enumerate}
\item There must be at least $4 R$ fields.
\item All $R$ fields may not exist in a single category. 
\item If $R = 4$, then only permutations of $(2_R, 2_R, n-4_{NS})$ are allowed. 
\item If $R > 4$, then no $NS$ are allowed in the maximal $R$ category (if one exists).  
\end{enumerate}
\end{enumerate}

\subsection{Non-Abelian Flat Directions and Self-Cancellation\label{NASC}}

In \cite{cfn2} we classified MSSM producing singlet field 
flat directions of the FNY model and in \cite{cfn3}
we studied the phenomenological features of these singlet directions.
Our past investigations suggested that for several phenomenological reasons,
including production of viable three generation quark and lepton mass matrices
and Higgs $h$-$\lh$ mixing, 
non-Abelian fields must also acquire FI-scale VEVs.  

In our prior investigations we 
generally demanded ``stringent'' flatness. 
That is, we forced each superpotential term to satisfy $F$-flatness
by assigning no VEV to at least two of the constituent fields.
While the absence of any non-zero terms from within $\vev{F_{\Phi_m}}$ and 
$\vev{W}$ is clearly sufficient to guarantee $F$-flatness along 
a given $D$-flat direction, 
such stringent demands are not necessary.
Total absence of these terms can be relaxed, so long as they appear in
collections which cancel among themselves in 
each $\vev{F_{\Phi_m}}$ and in $\vev{W}$. 
It is desirable to examine the mechanisms of such cancellations
as they can allow additional flexibility
for the tailoring of phenomenologically viable particle properties while
leaving SUSY inviolate. 
It should be noted that success along these lines may be short-lived,
with flatness retained in a given order only to be lost at one slightly higher.

Since Abelian $D$-flatness constraints limit only VEV magnitudes, we are left
with the gauge freedom of each group (phase freedom, in particular, is
ubiquitous) with which to attempt a cancellation between terms (whilst
retaining consistency with non-Abelian $D$-flatness).
However, it can often be the 
case that only a single term from $W$ becomes an offender in a given
$\vev{F_{\Phi_m}}$ (cf.\  Table 1B).  If a contraction of \NA fields (bearing
multiple field components) is present it may be possible to effect a
{\it self-cancellation} that is still, in some sense, ``stringently'' flat. 

Near the string scale the complete FNY gauge group is 
\beqn
&&[SU(3)_{C}\times SU(2)_{L}\times U(1)_{C}\times U(1)_{L}\times
U(1)_{A}\times \prod_{i=1'}^{5'} U(1)_i\times U(1)_4]_{\rm obs} \times
\nolabel\\
&&[SU(3)_H\times SU(2)_H\times SU(2)_{H'}
\times U(1)_{H}\times U(1)_{7} \times U(1)_{9}]_{\rm hid}\, .
\label{entgg}
\eeqn
The FNY non-Abelian hidden sector fields are triplets of $SU(3)_H$ or 
doublets of $SU(2)_H$ or $SU(2)_{H'}$. 
Self-cancellation of $F$-terms, that would otherwise break 
observable sector supersymmetry far above the electro-weak scale,
might be possible for flat directions containing such doublet or triplet 
VEVs. 
Since intermediate scale $SU(3)_H$ triplet/anti-triplet condensates 
are more likely to produce viable observable sector electro-weak scale 
supersymmetry breaking than are their $SU(2)_{H^{(')}}$ counterparts,  
we focus herein
on \NA directions containing doublet, but not triplet, FI-scale VEVs.

Whenever ``spinors'' of $SU(2)$ appear in $W$, they are not of the form
$S^{\dagger}S$, but rather are in the antisymmetric contraction
\beq
S_1 \cdot S_2 \equiv
S^T_1
\left (
\begin{array}{cc}
 0 & 1 \\
-1 & 0 \\
\end{array} \right )
S_2 \,\, .
\eeq
This form, which avoids complex conjugation and thus satisfies the requirement
of analyticity, is also rotationally (gauge) invariant as can be verified using
$\{\sigma_i , \sigma_j\} \nolinebreak[4] = \nolinebreak[4] 2 \delta_{ij}$,
$[\sigma_2 , \sigma_2] \nolinebreak[4] = \nolinebreak[4] 0$, 
and Eqs. (\ref{pauli}, \ref{rotmax})\, :
\beq
\sigma_2 R(\vec{\theta}) =
R^{\ast}(\vec{\theta}) \sigma_2
\eeq
\beq
S_1' \cdot S_2' =
(RS_1)^T (i \sigma_2) (RS_2) =
S_1^T (i \sigma_2) (R^{\dagger}R) S_2 =
S_1 \cdot S_2
\eeq
>From Eq. (\ref{spinor}), the general form of such a contraction may
be written explicitly as 
\beq
\label{contract}
S(\theta,\phi) \cdot S(\Theta,\Phi) =
- \sin(\frac{\theta - \Theta}{2})\cos(\frac{\phi - \Phi}{2})
-i \sin(\frac{\theta + \Theta}{2})\sin(\frac{\phi - \Phi}{2})
\,\, .
\eeq
The magnitude of this term must be a purely geometrical quantity and can be
calculated as
\beq
\vert S(\hat{n}) \cdot S(\hat{N}) \vert =
\sqrt{\frac{1 - \hat{n} \cdot \hat{N}}{2}} =
\sin(\frac{\delta}{2}) 
\,\, , 
\eeq
where $\delta (0 \rightarrow \pi)$ is the angle between $\hat{n}$ and $\hat{N}$.
The absence of a similar concise form for the phase is not a failing of
rotational invariance, but merely an artifact of the freedom we had in choosing
(\ref{spinor}).
Self-cancellation of this term is independent of the spinor's lengths and
demands only that their spatial orientations be parallel\footnote{The
contraction of a field with itself vanishes trivially.}. The same conclusion
is reached by noting that antisymmetrizing the equivalent (or proportional)
spinors yields a null value.  VEVs satisfying this condition are clearly not
$D$-consistent unless other \NA VEVed fields also exist such that the
{\it total} ``spin'' vector sum remains zero\footnote{In the notation of
\cite{cfn3}, taking a single sign for each of the $s_{k'}$ is a special case of
\NA self-cancellation, as is $\sum_{k=1}^p n_k s_k = 0$ a special case of the
$D$-constraint.}.  To examine generic cases of cancellation
between multiple terms, the full form of (\ref{contract}) is needed.

As an important special case, consider the example of a superpotential term
$\phi_1 \ldots \phi_n S_1 S_2 S_3 S_4$\footnote{Here and in the following
discussion we consider the doublets of a single symmetry group.} with
$\phi_n$ Abelian.  This is shorthand
for an expansion in the various pairings of non-Abelian fields, 
\beq
W \propto
\phi_1 \ldots \phi_n
\{(S_1 \cdot S_2)(S_3 \cdot S_4) +
(S_2 \cdot S_3)(S_1 \cdot S_4) +
(S_3 \cdot S_1)(S_2 \cdot S_4)\}\, ,
\label{na4}
\eeq
Broadly, we notice that:
\begin{itemize}
\item Whenever each term holds the same field set the spinors may be treated as
normalized to one, bringing any larger magnitudes out front as overall factors.
Furthermore, since $S^T$ appears but never $S^{\dagger}$, the same can
be done with any phase selections.
\item Since the contractions are antisymmetric, sensible interpretation of terms
with multiple factors demands the specification of an ordering.
\end{itemize}
The appropriate ordering, or equivalently the choice of relative signs, for
(\ref{na4}) is such to ensure {\it total} antisymmetrization.
When (\ref{na4}) is explicitly evaluated using the previously established
formalism it is seen to vanish identically for {\it all} field values.
The calculation is simplified without loss of generality by taking 
$\theta_1 = \phi_1 = \phi_2 = 0$.  We emphasize the distinction between
this identical exclusion from the superpotential and cancellations which exist
only at the vacuum expectation level.  $W$-terms with $6$ \NA fields are formed
with factors of (\ref{na4}) and also vanish, as do all higher order terms.

Even safe sectors of $W$ (in particular with $\vev{\Phi_{m}} = 0$) may yield
dangerous $\vev{F_{\Phi_{m}}} \nolinebreak[4] \equiv \nolinebreak[4]
\vev{\frac{\partial W}{\partial \Phi_{m}}}$
contributions.  The individual $F$-terms may be separated into two
classes based on whether or not $\Phi_{m}$ is Abelian.  For
the case of $\Phi_{m}$ non-Abelian, $\vev{F_{\Phi_{m}}}$ is itself a doublet.
As a note, terms like $\vev{F_{S_4}} \nolinebreak[4] \equiv \nolinebreak[4]
\vev{\frac{\partial W}{\partial S_4}}$ which {\it would have} arisen out of
(\ref{na4}) are cyclically ordered and also vanish identically.

\subsection{Cancellation in Terms Bearing $\p4$ VEVs\label{rizos}}

The four fields designated as `$(\p4)$' form a special system which merits
some focused discussion.  The pair $(\p4, \pp4)$ carry identical charges for all
$U(1)$'s and we must then consider a seperate term for each way to divide the
field content.  However, $\p4$ is from Neveu-Schwarz category two while
$\pp4$ is from category three and
picture-changing may disallow one or both fields from a term.
The same holds for their oppositely-charged vector partners.
Furthermore, it is phenomenologically indicated that providing FI-scale VEVs
to {\it all four} fields will decouple fractionally charged exotics to this
heavy mass scale.  This results, selection rules permitting, in infinite towers
of `new' potentially dangerous charge conserving terms
formed from each piece of $W$ by
adding $N$ of $(\p4, \pp4)$ and $N$ of $(\pb4, \ppb4)$.  $D$-flatness
(\ref{dtab}) cannot uniquely fix the VEV scales, but demands only that
\beqn
\mvev{\Phi_4} + \mvev{\Phi^{'}_4} - (\mvev{\bar{\Phi}_4} +
\mvev{\bar{\Phi}^{'}_4}) = a\times \mvev{\alpha}\, ,
\label{fiscale}
\eeqn
where $a$ is an integer (possibly negative) and $\vev{\alpha}$ is the
FI-scale for a given flat direction.  Similar effects occur whenever
the vector partner of any field takes on a VEV.  An additional constraint
\cite{cfn1} specific to the $(\p4)$ system is
\beqn
(\vev{\p4}\vev{\ppb4} + \vev{\pb4}\vev{\pp4}) = 0\, .
\label{constr}
\eeqn
$F$-flatness is destroyed by a tadpole term unless this is obeyed.

Using the rules stated in Section \ref{rizrules}, we will now systematically
discuss when picture-changing allows a $(\p4)$ tower to be added onto valid
existing terms in $W$.
\begin{description}
\item[(A)] If $n = 3$,
\\then there is no tower because no addition of purely Neveu-Schwarz fields can lead
to an acceptable term (with at least four Ramond fields).
\item[(B)] If $R = 4$, $(n-4_{NS}, 2_R, 2_R)$, with $R$ fields in categories
$2$ and $3$,
\\then there is no tower because $NS$ fields may not be added to
categories $2$ or $3$.
\item[(C)] If $R = 4$ with $R$ fields in category $1$ {\it or} $R > 4$ with an
$R$ maximum in $2$ or $3$,
\\then there is a tower but only ($\p4$, $\pb4$) or ($\pp4$, $\ppb4$) can
participate in it and just a single term exists at each order.
\item[(D)] If $R > 4$ and there is no maximal $R$
category {\it or} category $1$ is maximal,
\\then the full tower is present without restrictions.
\end{description}
If vector partners of other fields have VEVs then {\it each} term in their tower
is to be seen as a potential base point for its {\it own} $(\p4)$ tower.

The presence of all these similar terms begs an investigation of when a
self-cancellation is possible.  As caveats, we will
\begin{itemize}
\item only consider cancellation within a single ``base term'' of $W$, and
only within a single order at a time,
\item naively assume a uniform non-renormalizable coupling
for each term in a given cancellation.
\end{itemize}
Of the scenarios listed above, only (D) has multiple terms from $(\p4)$'s at a
given order and can sustain self-cancellation.  However, if the base order
$W$ term has no explicit $(\p4)$
factors then it cannot participate and the effort
is useless.  We will now undertake a detailed look at the more interesting
case of $m$ fields from ($\p4$, $\pp4$) in the base term of a tower.  The
discussion for ($\pb4$, $\ppb4$) follows directly.  In general, to effect a
cancellation at this lowest level, it is required that
\beqn
\sum_{i=0}^m \vev{\p4}^i \vev{\pp4}^{m-i} = 0\, .
\label{lev1}
\eeqn
At an order ($2N$) above the base order,
\beqn
&&\sum_{i=0}^{m+N} \sum_{j=0}^N [\vev{\p4}^i \vev{\pp4}^{(m+N)-i}]
[\vev{\pb4}^j \vev{\ppb4}^{N-j}] = 0
\\= &&\sum_{i=0}^{m+N} \sum_{j=0}^N [\vev{\p4}^i \vev{\pp4}^{m-i}]\vev{\pp4}^N
[\vev{\pb4}^j \vev{\ppb4}^{N-j}]\, .
\label{levN}
\eeqn
Factoring (\ref{lev1}) out of (\ref{levN}) and shifting the $i$ index, leaves
$\sum_{i=1}^N \sum_{j=0}^N [\vev{\p4}^{i+m} \vev{\pp4}^{N-i}]
[\vev{\pb4}^j \vev{\ppb4}^{N-j}]$, or for $N=1$,
\beqn
\vev{\p4}^{m+1} (\vev{\pb4} + \vev{\ppb4}) = 0\, .
\label{lev2}
\eeqn
To satisfy Eq. (\ref{lev1}), we must take {\it both} or {\it neither}
of ($\vev{\p4}$, $\vev{\pp4}$) as zero.
Taking neither leads to the restriction from (\ref{lev2}) that $\vev{\pb4}
= -\vev{\ppb4}$ and, unless we choose $\vev{\pb4} = 0$, this leads via
(\ref{constr}) to $\vev{\p4} = \vev{\pp4}$, which is clearly inconsistent
with (\ref{lev1}).  If we do take $\vev{\pb4} = 0$, then the expectation
value of the tower does vanish and the solution is consistent as
long as `$a$' from Eq. (\ref{fiscale}) is positive.  If $a$ equals zero,
the only
consistent solution is the trivial $\vev{\p4} \nolinebreak[4] = \nolinebreak[4]
\vev{\pp4} \nolinebreak[4] = \nolinebreak[4]
\vev{\pb4} \nolinebreak[4] = \nolinebreak[4]
\vev{\ppb4} \nolinebreak[4] = \nolinebreak[4] 0$.
If $a$ is negative, then ($\vev{\p4}$, $\vev{\pp4}$) must be zero and since all
terms in the tower contain these fields, its expectation value vanishes as well.
This discussion has been for towers added to $F$-dangerous terms only.
For the case of a non-vanishing contribution to $\vev{W}$ itself, derivatives
with respect to the $(\p4)$'s must be considered and ($\vev{\p4}$, $\vev{\pp4}$)
are now {\it demanded} to be zero (which is inconsistent for $a$ positive).

It seems that, at least within the context of our assumptions, only fairly
simple solutions exist which are at odds with the phenomenological imperative
to assign FI-scale VEVs to all four of the ($\p4$).  It also seems likely
that removing those assumptions would only serve to make cancellation more
difficult.

Finally, we will briefly take up the question of when and how the lowest level
term, (\ref{lev1}) itself, can be zero.  A natural choice seems to be
$\mvev{\p4} = \mvev{\pp4}$, although in this scenario the original
motivation for FI-scale
VEVs has been largely forsaken.  It might be that the resulting VEV$^2$s from
(\ref{fiscale}) are
fractional parts of $\mvev{\alpha}$, but this would be no more strange than
the integer portions currently assigned to other fields.
Putting $\vev{\p4} = e^{i\theta}\vev{V}$ and $\vev{\pp4} = \vev{V}$
into Eq. (\ref{lev1}) yields
\beqn
\vev{V}^m \sum_{\gamma=0}^m e^{i\gamma\theta} = 0\, .
\eeqn
`Tip-to-tail' solutions in the complex plane are always possible,
for example, with $\theta = \frac{2\pi}{m+1}$\, .
However, it must be kept in mind that a single set of VEVs for the $(\p4)$ must
be chosen {\it simultaneously} for the entire `flat direction' and there may be
issues of compatibility between the various incarnations of Eq. (\ref{lev1}).

\section{Minimal Standard Heterotic-String Model Non-Abelian Flat Directions\label{mssmfds}} 

Our initial systematic search for MSSM-producing stringent flat directions 
revealed four singlet directions that were flat to all order, 
one singlet direction flat to twelfth order, and numerous singlet 
directions flat only to seventh order or lower \cite{cfn2}. 
For these directions renormalizable mass terms appeared for one complete
set of up-, down-, and electron-like fields and their conjugates.
However, the apparent top and bottom quarks did not appear in the 
same $SU(2)_L$ doublet. Effectively, these flat directions gave the
strange quark a heavier mass than the bottom quark. This inverted 
mass effect was a result of the field $\p{12}$ receiving a VEV
in all of the above direction. 

We thus performed a search for MSSM-producing singlet flat directions
that did not contain $\vev{\p{12}}$. None were found. 
This, in and of itself, suggests the need for 
non-Abelian VEVs in more phenomenologically
appealing flat directions. 
Too few first and second generation down and electron mass terms
implied similarly. 

One interesting aspect of the singlet flat direction mass matrices
was the implication that the vector partner of $\p{12}$ should 
take on a VEV, but both renormalizable flatness constraints and 
viable Higgs $\mu$-terms required $\vev{\pb{12}}$ to be many orders
of magnitude below the FI VEV scale of the flat direction fields.
This possibility will be reconsidered in our three generation  
mass matrix discussions in Section 5. 

\setcounter{footnote}{0}
The results of our non-Abelian flat direction search, which we
summarize in Tables 1A and 1B, proved interesting.
`FDNA\#' designations from \cite{cfn4} are continued.
In general, the number of distinct $U(1)$ charges in a model is equal
to the minimum number of fields, charged {\it independently} under
those $U(1)$'s, which must take
VEVs in order for dangerous gauge invariant terms to emerge in $\vev{W}$.
In fact, FDNA5, which has only nine fields taking VEVs (counting $(\p4)$ once),
achieves its $F$-flatness in just this way, unable to obey gauge constraints
without two or more unVEVed fields\footnote{The third order $(\p4)$ term
mentioned near Eq. (\ref{constr}) will appear, but it is cancelled.}.
However, $D$-flat directions of this sort will be very limited.
Equation (\ref{dtab}), which does not restrict the anomalous charge and
thus imposes one fewer condition than $W$, allows maximally\footnote{Although
an {\it algebraic} gauge invariant solution can always be generated, the
restriction to integer field number coefficients and signs which respect
unassigned vector partner VEVS, can greatly reduce the incidence of
{\it physical} solutions from the start.\label{fnalg}} a single solution
at this level per set of VEVed fields.
Alternatively, if you are ``one field short'', the inclusion of a field with no
VEV may$\ref{fnalg}$ produce one gauge invariant $F$-term.
The remaining directions listed in Table 1A, besides FDNA(5+8), denoting
the specific combination 18 FDNA5 + 1 FDNA8, are of this
sort.  They contain ten fields with VEVs and are dangerous to SUSY only
through isolated $\vev{F}$ terms, but never $\vev{W}$.
Since there are a finite number of fields which produce $\vev{F_{\varphi_i}}$
contributions, if all possibilities have been exhausted and the direction
remains safe, the result holds to all orders.

We discovered several MSSM-producing $D$-flat directions that
are stringently $F$-flat to at least seventh order
and that satisfy the top-bottom $SU(2)$ doublet requirement of
$\vev{\p{12}}= 0$. Seven of these directions are $F$-flat to all
finite order.  The simplest, FDNA5, is the only direction for which
no $F$-breaking terms appear in a gauge invariant superpotential.
For both FDNA8 and FDNA9, there is just one respective
dangerous (eleventh order) gauge invariant term and it is not invariant under
the picture-changed global worldsheet symmetries of the string theory\footnote{
This can be shown \cite{rizospc} based on the rules for allowed picture-changed
global worldsheet charge combinations presented in \cite{nansr,rizos91} and
summarized in section \ref{rizrules}.}.  Thus, neither term will appear in the
low energy effective field theory superpotential.  For FDNA6, one twelfth,
two thirteenth, one seventeenth, and one eigthteenth order dangerous gauge
invariant terms do exist.  Again, stringy worldsheet charge requirements
eliminate terms, but this time one of the thirteenth order offenders remains.
However, it is of the form of Eq. (\ref{na4}), containing four \NA fields, and
vanishes identically.  Likewise, FDNA7, FDNA10, and FDNA17 are flat to all order
thanks to the selection rules and excessive \NA field content.

The remaining VEV directions listed in Table 1A are broken at finite order.
FDNA11 and FDNA12 contain dangerous ninth order terms.  The {\it same} eighth
order term spoils each of FDNA13, FDNA14, FDNA15, and FDNA16.  FDNA18
$F$-flatness breaks at seventh order.  Interestingly, in all of these cases
there is only ever a single term still standing, but for none of them
do $D$-flatness requirements permit a consistent \NA self-cancellation
solution.

We also include FDNA(5+8) in the tables, which breaks $F$-flatness at only fifth order,
but has none the less provided an interesting study.
In particular, we have found terms ranging from fourth to tenth order
which {\it do} experience a self-cancellation.  It will be discussed shortly
in much greater detail.

\section{Flat Direction Phenomenology}

\subsection{Higgs Mass Matrix and $\mu$-Terms\label{handmu}}

The FNY model contains four fields, $\h{1,2,3}$ and $\h{4}\equiv \H{41}$,
that can play the role of the MSSM Higgs doublet $h$ 
and four corresponding fields, $\bh{1,2,3}$ and $\bh{4}\equiv \H{34}$,
that can play the role of the Higgs doublet $\lh$.
Higgs mass ``$\mu$-terms'' take the form of one field from each of these
sets plus one or more fields which take a VEV.
Collectively, they may be expressed in matrix form as the scalar contraction
\beqn
\h{i} M_{ij} \bh{j}\, .
\label{hmm}
\eeqn
Specifically, we are interested in developing unique massless
eigenstates $h$ and $\bar h$ formed from linear combinations of the $\h{i}$
and $\bh{i}$ respectively,\footnote{The possibility of linear  
combinations of MSSM doublets forming the physical Higgs is a feature 
generic to realistic free fermionic models.}  
\beqn
  h = \frac{1}{n_h}    \sum_{i=1}^{4} c_i h_i;
\quad\quad
\lh = \frac{1}{n_{\lh}} \sum_{i=1}^{4} \bc_i \bh{i}\, ,
\label{bhdef}
\eeqn
with normalization factors $n_h = {\sqrt{\sum_i (c_i)^2}}$,
and  $n_{\lh} = {\sqrt{\sum_i (\bc_i)^2}}$.  These combinations
will then in turn establish the quark and lepton mass matrices.

However, the physical mass terms are those appearing in the scalar Lagrangian.
They do not exist in the superpotential itself, but rather arise out of it in
the form $| \frac{\partial W}{\partial \Phi_{m}} |^2$.  Thus, the mass terms
relevant to the fields ($\bh{j}$) are produced from derivatives of Eq.
(\ref{hmm}) with respect to the $h_i$, and may be written as
\beqn
\sum_{i=1}^{4} | M_{ij} \bh{j} |^2 = (\lh)^{\dagger}M^{\dagger}M(\lh)\, .
\label{laghiggB}
\eeqn
The matrix $M^{\dagger}M$ will then yield the $m^2_{\lh}$ eigenvalues
when diagonalized.  Since the matrix is Hermitian these will be real, although they
do remain sensitive to phase selections within $M$.  Furthermore, since the
eigenvectors are orthogonal (or can be chosen so with degeneracy), they may be
used, once normalized, to construct a unitary matrix
$U \equiv
\left (
\begin{array}{ccc}
u_1 & u_2 & \cdots \\
\downarrow & \downarrow &
\end{array} \right )$.
The diagonalization of $M^{\dagger}M$ proceeds by inserting factors\footnote{
The two instances of $(UU^{\dagger})$ must be identically composed if the
defining relation, $(U^{\dagger}MU) \nolinebreak[4] (U^{\dagger} \vert \Lambda
\rangle) \nolinebreak[4] = \nolinebreak[4] \lambda (U^{\dagger} \vert \Lambda
\rangle)$, is to hold for the diagonal basis.} of $\Io = (UU^{\dagger})$ into
(\ref{laghiggB}), ${(\lh)}^{\dagger}UU^{\dagger}M^{\dagger}MUU^{\dagger}(\lh)$,
so as to facilitate a grouping into four distinct mass values without disrupting
the actual collection of terms.  The right-most matrix,
$U^{\dagger} \equiv
\left (
\begin{array}{cc}
{u_1}^{\ast} & \rightarrow \\
{u_2}^{\ast} & \rightarrow \\
\cdots &
\end{array} \right )$,
serves to receive column vectors and project out their diagonal coefficients,
while the adjacent factor of $U$ ensures that vectors from the diagonal basis
properly interpolate to the original\footnote{
This is a statement of the completeness relation,
$\sum \vert \Lambda \rangle \langle \Lambda \vert = \Io$}
before passing through $M^{\dagger}M$.
Similarly, the $(h_i)$ mass terms are
\beqn
\sum_{j=1}^{4} | h_i M_{ij} |^2 = (h)^TMM^{\dagger}{(h)}^{\ast}\, .
\label{laghiggs}
\eeqn
The $MM^{\dagger}$ eigenvectors, denoted as $(v_i)$, compose a matrix we will
call $U^{'}$.  Finally, the massless physical Higgs doublets may be expressed as
\beqn
h   = \sum_{i=1}^{4} {v}^{(0)}_i h_i;
\quad\quad
\lh = \sum_{i=1}^{4} {u^{\ast}}^{(0)}_i \bh{i}\, ,
\label{physhigg}
\eeqn
the zero-mass elements of $({U^{'}}^Th)$ and $(U^{\dagger}\lh)$
respectively.

We have found that both singlet and non-Abelian MSSM-producing 
flat directions 
necessarily contain $\p{23}$, $\sH{31}$, and $\sH{38}$ VEVs. 
Together these three VEVs produce four terms in the Higgs
mass matrix: $\h{3}\bh{2} \vev{\p{23}}$, $\h{2}\bh{4} \vev{\H{31}}$,  
$\h{4}\bh{3} \vev{\H{38}}$, and $\h{4}\bh{4} \vev{\H{31}}$.
When these are the only non-zero terms in the matrix, 
the massless Higgs eigenstates
simply correspond to $c_1 = \bc_1 = 1$ and  $c_j = \bc_j = 0 $ for $j=2,3,4$.  
In this case all possible quark and lepton mass terms of the form 
$\Q{m} \uc{n} \bh{j}$, $\Q{m} \dc{n} \h{j}$, $\Q{m} \ec{n} \h{j}$,
$\L{m} \Nc{n} \h{j}$, $\L{m} \L{n} \h{i} \h{j}$, where
$m,n\in \{ 1,2,3\}$, $j\in \{2,3,4\}$, and $i\in \{1,2,3,4\}$ decouple 
from the low energy MSSM effective field theory. However, when one
or more of the $c_j$ or $\bc_j$ are non-zero, then some of these terms
are not excluded and provide addition quark and lepton mass terms.
In such terms, the Higgs components can be replaced by their corresponding
Higgs eigenstates along with a weight factor,
\beqn
h_i   \rightarrow \frac{c_i}{n_h} h;
\quad\quad 
\lh_i \rightarrow \frac{\bc_i}{n_{\lh}} \lh\, . \label{bhir}
\eeqn 
Thus, in string models such as the FNY, two effects can contribute
to inter-generational (and intra-generational) mass hierarchies:
generic suppression factors of $\frac{\vev{\phi}}{\MP}$ in 
non-renormalizable effective mass terms and 
$\frac{c_i}{n_h}$ or $\frac{\bc_i}{n_{\lh}}$ 
suppression factors. This means a hierarchy of values among the 
${c_i}$
and/or among the ${\bc_i}$ holds the possibility of producing
viable inter-generational $m_t:m_c:m_u = 1: 7\times 10^{-3}: 3\times 10^{-5}$ mass ratios even when 
all of the quark and lepton mass terms are of renormalizable or very 
low non-renormalizable order order. 
Note that more than one generation
of such low order terms {\it necessitates} a hierarchy among the 
$c_i$ and $\bc_i$.    
Generational hierarchy via suppression factor in Higgs components was first
used in free fermionic models of the flipped $SU(5)$ class, see for example \cite{higsup}.

In Appendix B, we have provided all possible quark and electron-like
mass terms  through eighth order, for which the fourth through last
fields in these terms must acquire a VEV. 
The FNY up- and down-like renormalizable terms are 
\beqn
\bh{1} \Q{1 }  \uc{1}\, ,\quad
\h{2}  \Q{2 }  \dc{2}\, ,\quad  
\h{3}  \Q{3 }  \dc{3}\, .
\label{udrenorm}
\eeqn
Since $\bh{1}$ is either the only component or a primary component in
$\lh$, the top quark is necessarily contained in $\Q{1}$ and $\uc{1}$
is the primary component of the left-handed anti-top mass eigenstate.
Thus, as we have already discussed, the bottom quark must be the second
component of $\Q{1}$. 
Since there are no renormalizable 
$\h{i} \Q{1} \dc{m}$ terms in Eq.~(\ref{udrenorm}),
a bottom quark mass that is  hierarchically
larger than the strange and down quark masses requires that 
\beqn
\frac{|c_{j=2,3}|}{n_h}\ll 1\, .
\label{c23h}
\eeqn
Non-zero $c_{j=2,3}$ satisfying Eq.\ (\ref{c23h}) could,
perhaps, yield viable strange or down mass terms.  

The first possible bottom mass term appears at fourth order, 
\beqn
\h{4}\Q{1 }\dc{3}\sH{21}\, .
\label{bot4}
\eeqn
Realization of the bottom mass via this term would require
$h$ to contain a component of $\h{4}$ and for
$\sH{21}$ to acquire a VEV. Of our flat directions, only FDNA8
and FDNA(5+8) give $\sH{21}$ a VEV. Of these two, only
FDNA(5+8) embeds part of $\h{4}$ in $h$.  

The physical mass ratio of the top and bottom quark is
of order $\sim 3\times 10^{-2}$. 
In free fermionic models there is apparently 
no significant suppression to the numeric value of an 
effective third order superpotential coupling constant,
$\lambda_3^{\rm eff} = \lambda_4 \vev{\phi}$,
originating from a fourth order term. 
Hence, a reasonable top to bottom mass ratio would imply 
\beqn
\frac{|c_2|}{n_h},{|c_3|}{n_h} \ll 
\frac{|c_4|}{n_h}\sim 10^{-2\,\,\, {\rm to}\,\,\, -3}
\label{c4ran}
\eeqn
when $\frac{|\bc_1|}{n_{\lh}}\sim 1$ and $\vev{h}\sim \vev{\bar{h}}$.
 
The next possible higher order bottom mass terms do not occur 
until sixth order:
\beqn 
{\begin{tabular}{lll}
 $\pps\h{2}\Q{1 }\dc{3}\p{13} \sV{2} \sV{21}$   
&$+   \h{2}\Q{1 }\dc{3}\p{13}  \V{9}  \V{29}$ 
&$+   \h{3}\Q{1 }\dc{2}\p{12} \sV{2} \sV{11}$\\  
 $+   \h{3}\Q{1 }\dc{2}\p{12}  \V{5}  \V{17}$  
&$+   \h{4}\Q{1 }\dc{3}\sH{15}\sH{18}\sH{19}$\, .  
\label{bot6}
\end{tabular}\nolabel}
\eeqn
Beyond fourth order a suppression factor of 
$\frac{1}{10}$ per order is generally assumed.
Thus, a sixth order down mass term such as these would imply
$\frac{|c_j|}{n_h} \sim 1$, where  $j$ is one of  $\{2,3,4\}$ as
appropriate, when $\frac{|\bc_1|}{n_{\lh}}\sim 1$. 
However, none of our flat directions 
have sufficient VEVs to turn any of the \eq{bot6} terms into mass terms.

If not sixth order, then seventh order is probably the
highest order that could provide a sufficiently large bottom mass. 
There are no such seventh order terms containing $\h{1}$. However, 
$\h{2}$ is in 15 of these terms, 
of which 
\beqn 
\h{2}\Q{1 }\dc{2}\Nc{3} \sH{31} \H{26} \V{37}
\label{bot7a}
\eeqn
becomes a bottom mass term
for FDNA(5+8) and of which
\beqn 
\h{2}\Q{1 }\dc{3}\Nc{2} \sH{31} \H{26} \V{37} 
\label{bot7b}
\eeqn
becomes a bottom mass term
for FDNA7 (and for additional directions that lose flatness at tenth
order or lower).
$\h{3}$ is in 3 seventh order terms, but none of these become mass terms 
for any of our flat directions. 
$\h{4}$ appears in 17 of these terms. $\p{12}$, however, appears in 15 of 
the 17 and is forbidden a VEV. $\p{13}$, which also doesn't acquire a
VEV in any of our directions, appears in the remaining two.   

Therefore, 
the only possible bottom quark mass terms 
resulting from our set of flat directions under investigation, 
are the fourth order term (\ref{bot4}) involving $\h{4}$, and 
the seventh order terms (\ref{bot7a}) and
(\ref{bot7b}) involving $\h{2}$.   
Consider first the fourth order term. For this we must ask if
a $\frac{|c_4|}{n_h}\gsim 10^{-3}$ value can be realized 
in our FNY model. 
This would require either a $\h{1}\bh{3}$ or $\h{1}{\bh4}$ mass term. 
While both terms result in a non-zero $|c_4|$,
one can be shown that a $\h{1}\bh{3}$ term gives $|c_4| \gsim  |c_2|$, 
while a $\h{1}\bh{4}$ term results in $|c_4| \lsim  |c_2|$. 
Therefore, we seek a $\h{1}\bh{3}$ term. 
Below seventh order, 
the only such term is $\h{1}\bh{3}\p{13}$.
$\frac{|c_4|}{n_h}\gsim 10^{-3}$ would require
$\vev{\p{13}} \gsim 10^{-3}$ FI-scale. 
However, the trilinear term $\bp{12}\p{13}\p{23}$ forbids
$\bp{12}$ or $\p{13}$ from taking on, with $\p{23}$, a near 
FI-scale VEV along a stringently $F$-flat direction.

Thus, we must consider seventh or higher order terms
that could give $\h{1}\bh{3}$ mass terms.
In Appendix B, we provide the complete set of $\h{1}\bh{3}$ mass 
terms through eighth order.
We find that none of our Table 1A flat directions contain
appropriate VEVs to transform any of the seventh or eighth
order terms into
effective mass terms. 
While several flat directions can generate a particular ninth order 
mass term,
\beqn
\h{1}\bh{3}\vev{\Nc{3} \pp{4}\sH{15}\sH{30}\sH{31}\H{28}\cdot\V{37}}\, .
\label{h1bh39a}
\eeqn
FDNA(5+8) is the only one of these that 
simultaneously generates the fourth order bottom mass term.  FDNA(5+8) has
been singled out for more detailed examples and analysis in the following
sections.  It produces good opportunities for study of cancellations as
well as phenomenologically interesting near-string-scale quark, lepton,
and Higgs mass matrices.  However, it must be considered as purely a training
exercise since the two terms which escape all of our filters break supersymmetry
at only fifth order.

\subsection{Flat Direction (5+8)\label{58FD}}

The VEV set designated as FDNA(5+8) is formed (cf. Table 1A of Appendix A)
as the linear combination 18 FDNA5 + 1 FDNA8.  More specifically, the
absolute values squared of the FI-scale coeffients are
combined in this ratio.  Hybrid directions with
desired properties may be ``engineered'' in this manner without disrupting 
Abelian $D$-flatness (\ref{dtab}).  Of course, the flaws of the constituent
VEV sets can be embedded into the result as well.
By using the all-order flat directions 5 and 8, this concern is alleviated.
The assigned factor of `18' brings all of the \NA VEVs to the same scale.

FDNA(5+8) contains fourteen fields and generates four linearly independent
basis vectors for the construction of gauge invariant terms in $\vev{W}$ alone.
In addition, each field without a VEV forms its own basis vector to be combined
with the original four into dangerous $F$-terms.  The result is a greatly
expanded number of terms which can break supersymmetry, relative to the simpler
directions of section (\ref{mssmfds}), but along with this
complexity can come a richer mass phenomenology and more interesting
opportunities for cancellation.  Since multiple terms may now appear within
each $F_{\Phi_{m}}$ (cf. \ref{ftgen}), one possibility  which emerges is a
cancellation among those components.  However, this seems to require excessive
tuning and to be unsustainable across many orders.

The basis combination was performed with a program
which covered a coefficient parameter space large enough to produce all terms
at or below order twenty five.  We searched for terms with integer numbers of
all fields and an even number of \NA fields\footnote{The VEVed \NA fields in
FDNA(5+8) are only from $SU(2)_{H'}$.} with either 0 or 1 unVEVed fields (i.e.\ 
terms existing in $W$ which FDNA(5+8) makes dangerous to $W$- or
$F$-flatness).  By choosing a basis set with at least one unique field per
component, such computations are greatly simplified and easily guaranteed
complete to a specified order.
This is because the possible coefficents become restricted
to a simple subset of rational numbers or, if the unique field's strength is 
divided to unity, integers.  Also, unless both the unique field and its vector
partner take a VEV, coefficients of only one sign need be considered to find
$F$-terms.
As with the other directions in Table 1A, this occurs here only for $(\p4)$.

Our search yielded 131 dangerous terms, five of them to $\vev{W}$ with
${11}^{th}$ the lowest order and 126 of them to $\vev{F}$, as low as order
four (counting variations of $(\p4)$ only once).  World sheet selection rules
reduced this number to 32, all of them $F$-terms.  Disallowing more than two
\NA fields (foreach $SU(2)$ group)
trimmed the list further to just the eight terms in
Table \ref{danger8}.  If a single incidence of $(\p4)$ is mandated, then
it is so indicated by a lack of parenthesis.

\beqn{\begin{tabular}{l|l|l}
\#& {\cal{O}}(W) & $F$-term \\ \hline\hline
1 &  4 & $H^s_{16}\, \vev{\H{26} \cdot \V{37}} \vev{N^c_3}$ \\ \hline
2 &  5 & $V^s_{32}\, \vev{\H{26} \cdot \V{37}} \vev{\p{4} \sH{37}}$
	\\ \hline
3 &  5 & $V_{15}\, \vev{\cdot \V{35}} \vev{\pp{4} \sH{30} \sH{21}}$ \\ \hline
4 &  5 & $V_{17}\, \vev{\cdot \V{5}} \vev {\pp{4} \sH{30} \sH{15}}$ \\ \hline
5 &  8 & $\bp{13}\,  \vev{\H{26} \cdot \V{37}} \vev{(\p{4}) \Hs{31} \sH{30}
	\sH{15} N^c_3}$ \\ \hline
6 &  9 & $\bp{13}\,  \vev{\V{5} \cdot \V{35}} \vev{\p{23} {(\p{4})}^2
	{\sH{30}}^2 \sH{21} \sH{15}}$ \\ \hline
7 &  9 & $\bp{12}\,  \vev{\H{26} \cdot \V{37}} \vev{\p{23} (\p{4}) \Hs{31}
	\sH{30} \sH{15} N^c_3}$ \\ \hline
8 & 10 & $\sH{36}\,  \vev{\H{26} \cdot \V{37}} \vev{\p{23} \p{4} \Hs{31}
	\sH{30} \sH{15} \sH{37} N^c_3}$
\label{danger8}
\end{tabular}}
\eeqn

The lowest order term (designated as \#1) contains a factor of $\vev{\H{26}
\cdot \V{37}}$ which we would like to cancel, as per the discussion of Section
\ref{NASC}.  This requires the VEV orientations to be chosen parallel
in the three-dimensional ${SU(2)}_H$ adjoint space.
Since FDNA(5+8) contains (two) additional \NA fields with VEVs ($\V{5}$ and
$\V{35}$) which can oppose $\H{26}$ and $\V{37}$ with an equal total magnitude,
this choice is also $D$-consistent.  The same factor appears in
and eliminates terms \#2,5,7 and 8.  Since the other two \NA VEVs had to be
parallel as well, the contraction $\vev{\V{5} \cdot \V{35}}$ in term \#6 is
also zero.  In the language of \cite{cfn3}, we could have said
$s_{\H{26}} = s_{\V{37}} = 1$ and $s_{\V{5}} = s_{\V{35}} = -1$.
This leaves us with only \#3 and \#4, both of which are fifth order
terms with unVEVed {\it \NA} fields so that self-cancellation is impossible.
Furthermore, they will appear in different $F$-terms and each allows only
a single $(\p4)$ configuration, ruling out a couple of other (less satisfactory)
scenarios.  The choice $\vev{\pp4} = 0$, along with $\vev{\ppb4} = 0$ for 
consistency with Eqs. (\ref{fiscale}, \ref{constr}), would restore
$F$-flatness by simply removing the offending terms from $\vev{F}$.
However,  as has
been discussed, this seems phenomenologically unviable and so it appears that
we are stuck with a broken FDNA(5+8) at order five.  As a note, the
cancellations which were sucessful are insensitive to the factor of 18 between
flat directions five and eight.  Also, while it is
common to see the vanishing of terms with excessive \NA doublets, these
mark the {\it only} examples wherein \NA self-cancellation by selected VEVs
have been found for the Table 1A flat directions.


\subsection{Higgs Mass Matrices for FDNA(5+8)}

Table 3 in Appendix C lists the up, down, and electron and Higgs mass 
matrix terms through order nine for FDNA(5+8). The Higgs terms in this table
produce the mass matrix
\beqn
M_{h_i,\hb{j}}&=&{\left ( 
\begin{array}{cccc}
0 & 0 & \lambda_9 \vev{X_{13}} &  0 \\
\vev{\pb{12}} & 0 & 0  & g \vev{\H{31}} \\ 
0 & g \vev{\p{23}} & 0  &  0 \\
0 & 0 & g\vev{H_{38}} & 
\lambda_5 \vev{X_{44}}
\end{array} \right )}
\label{mija}
\eeqn
where
$X_{13} = \frac{\Nc{3}\pp{4}\sH{15}\sH{30}\sH{31}\H{28}\cdot\V{37}}{\MS^6}$ and
$X_{44} = \frac{\phi_{23} \sH{31} \sH{38}}{\MS^2}$.

In \cite{cfn3,cfn4} we showed that a small $\pb{12}\ll$ FI-scale VEV 
produces superior quark and lepton mass matrix phenomenology.
Thus, in the matrix (\ref{mija}) we allow possible contributions from the mass term
$\h{2}\hb{1}\vev{\pb{12}}$.  However, since $\pb{12}$ does not acquire a
VEV in any of our flat directions in Table 1A, 
$\vev{\pb{12}}\ne 0$ would have to arise as a second order effect at or below the 
GUT scale. 

When we allow $\vev{\pb{12}}\sim 10^{-4}$,
the numeric form of the FDNA(5+8) Higgs doublet mass 
matrix is\footnote{As stated in Section \ref{handmu}, we will assume a
suppression in the superpotential coupling of $\frac{1}{10}$ per order
above fourth.}
\beqn
M_{h_i,\hb{j}}&\sim&{\left ( 
\begin{array}{cccc}
0 & 0 & 10^{-5} &  0 \\
10^{-4} & 0 & 0  & 1 \\ 
0 & 2 & 0  &  0 \\
0 & 0 & 1  & 10^{-1} 
\end{array} \right ).}
\label{mijb}
\eeqn
Using (\ref{physhigg}), the near-massless Higgs eigenstates
produced from \eq{mijb} are: 
\beqn
 h  &=& \h{1}  +  10^{-7} \h{2}  - 10^{-5} \h{4} \label{hes}\\
\lh &=& \bh{1} +  10^{-6} \bh{3} - 10^{-4} \bh{4}\, .\label{bhes}
\eeqn
Note that the coefficient of $\bh{4}$ is approximately $\pb{12}$. 
In the limit of $\vev{\pb{12}}=0$, the $h$ eigenstate reduces to $\bh{1}$.

These Higgs eigenstates provide examples of how 
several orders of magnitude mass suppression factors can appear 
in the low order terms of a Minimal Standard Heterotic-String Model.
Here, specifically, the $\h{4}$ coefficient in (\ref{hes}) can provide 
$10^{-5}$ mass suppression for one down-quark generation and one electron generation,
since $\h{4} (\Q{i} \dc{j} + \L{i} \ec{j})$, when rewritten in terms
of the Higgs mass eigenstates, contains a factor of  
$10^{-5} h (\Q{i} \dc{j} + \L{i} \ec{j})$. 
Similarly,             
the $\h{2}$ coefficient can provide $10^{-7}$ suppression. 
Further, the $\bh{4}$ coefficient in (\ref{bhes}) can provide 
$10^{-4}$ up like-quark mass suppression and            
the $\bh{3}$ coefficient a corresponding $10^{-6}$ suppression.
We examine the quark and charged-lepton eigenstates in the following 
subsection.

\subsection{Quark and Lepton Mass Matrices for FDNA(5+8)}
  
The matrices in this section are calculated up to ninth order, where the
suppression in the coupling (assumed here to be $\sim {10}^{-5}$) is comparable
to that coming out of Eqs. (\ref{hes}, \ref{bhes}).  The terms comprising these
matrices (cf. Appendix C, Table 3) take the form $\bh{i}\Q{j}\uc{k}$,
$\h{i}\Q{j}\dc{k}$, and $\h{i}\L{j}\ec{k}$, plus possibly some number of VEVed
fields.  Not having an exact value for the couplings, we will also not concern 
ourselves here with the specific magnitudes of the VEVS or overall phases for
the massterms.

The up-quark mass matrix for FDNA(5+8) contains only a single term
(corresponding to the top mass) when $\vev{\pb{12}}=0$, but develops an
interesting texture when $\vev{\pb{12}} \sim 10^{-4}$ FI-scale.
To leading order, the general form is  
\beqn
\nonumber
M_{\Q{},\uc{}}&=&{\left ( 
\begin{array}{ccc}
\hb{1} & .1 \hb{3}+ 10^{-3} \hb{4} & 0 \\
10^{-2}  \hb{3}+ 10^{-4} \hb{4} & 0 & 0 \\ 
0 & 0 & \hb{4} \\
\end{array} \right ).}
\label{muij}\\
             &\sim&{\left ( 
\begin{array}{ccc}
1 & 10^{-6} & 0 \\
10^{-7} & 0 & 0 \\ 
0 & 0 & 10^{-4} \\
\end{array} \right ).}
\label{muijn}
\eeqn
Just as for the Higgs, discussed in the start of Section \ref{handmu}, the
$m^2_{\uc{}}$ are obtained by diagonalizing $M^{\dagger}M$.
In top quark mass units, the up-like mass eigenvalues are $1$, $10^{-4}$, and $10^{-13}$. 
Clearly, a more realistic structure would appear if the $\hb{4}$ suppression 
factor was $10^{-2}$ rather than $10^{-4}$.   

The FDNA(5+8) down-quark mass matrix is independent of the $\pb{12}$ VEV value.
It has the form
\beqn
\nonumber
M_{\Q{},\dc{}}&=&{\left ( 
\begin{array}{ccc}
0 & 10^{-3}\h{2}+ 10^{-5} \h{4} & \h{4}\\
10^{-5} \h{2} & \h{2}+ .1 \h{4} & 10^{-5} \h{2} \\ 
0 & 10^{-4} \h{2}  & 0\\
\end{array} \right ).}
\label{mdij}\\
             &\sim&{\left ( 
\begin{array}{ccc}
0 & 10^{-9} & 10^{-5}  \\
10^{-11} & 10^{-6}  & 10^{-11} \\ 
0 & 10^{-10}  & 0 \\
\end{array} \right ).}
\label{mdijn}
\eeqn
The resulting down-quark mass eigenvalues   
are $10^{-5}$, $10^{-6}$, and $10^{-15}$ 
(again in top quark mass units).
One might suggest this provides quasi-realistic
down and strange masses, but lacks a bottom
mass. Unfortunately, the down-like quark eigenstate corresponding
to a $10^{-5}$ mass is in $\Q{1}$, arising with its suppression factor from
the term $\h{4} \Q{1} \dc{3} \sH{21}$.
From our discussion above, this is in fact the bottom quark.
The second and third generation masses would be more viable if
the $\h{4}$ suppression factor were $10^{-2}$ instead.    

The FDNA(5+8) charged-lepton mass matrix (likewise 
independent of the $\pb{12}$ VEV value) takes the form
\beqn
\nonumber
M_{\L{},\ec{}}&=&{\left ( 
\begin{array}{ccc}
0 & 10^{-4}\h{2} & 0\\
10^{-4} \h{2} & \h{2}+ .1 \h{4} & 10^{-4} \h{2} \\ 
.1 \h{4} & 10^{-5} \h{2}  & 0\\
\end{array} \right ).}
\label{meij}\\
             &\sim&{\left ( 
\begin{array}{ccc}
0 & 10^{-10} & 0  \\
10^{-10} & 10^{-6}  & 10^{-10} \\ 
10^{-5} & 10^{-11}  & 0 \\
\end{array} \right ).}
\label{meijn}
\eeqn
The three corresponding electron-like 
mass eigenvalues $10^{-5}$, $10^{-6}$,
and $10^{-14}$.  
As with the down-like quark masses, more viable 
second and third generation electron masses would appear
if the $\h{4}$ suppression factor was only $10^{-2}$.    

For both $M_{\Q{},\dc{}}$ and $M_{\L{},\ec{}}$, the $(2,2)$ element is composed
by two terms of similar magnitude.  The mass phenomenology degrades further
if these contributions cancel significantly.  Also, the cancellations considered
in Section \ref{58FD} are not without an effect on these mass ratios.  If
\NA self-cancellation is implemented, then the ninth order contribution to
$M_{h_i,\hb{j}}$, Eq. (\ref{mija}), vanishes and the massless Higgs
becomes simply $h_1$.  Since neither $\dc{}$ nor $\ec{}$ appear to this order
in mass terms with $h_1$, their mass hierarchies also vanish.  $\lh$ is 
insensitive to this particular change, as are the leading order terms of
$M_{\Q{},\uc{}}$, Eq (\ref{muijn}).  If $\vev{\pp4} = \vev{\ppb4} = 0$ were
further enforced, the only effect here would be the loss of the lightest
up-like quark.

\section{Concluding Remarks} 

The more realistic free fermionic string models are the most realistic
string models constructed to date.
By expanding our flat direction search to allow VEVs to non-Abelian
charged fields,\cite{cfn4} rather than limiting VEVs to non-Abelian 
singlets,\cite{cfn1,cfn2,cfn3}
we have improved the phenomenology of
our Minimal Standard Heterotic String Model. We have found that
quasi-realistic patterns to quark and charged-lepton mass matrices
can appear. 
The improved phenomenology was a result of both the non-Abelian VEVs
and the struture of the physical Higgs doublets $h$ and $\bar{h}$.
In the more realistic free fermionic heterotic models, each of 
these Higgs can contain up to four components with vastly different
weights, differing by several orders of magnitude. These components
generically all couple differently to a given $\Q{i}\uc{j}$,
$\Q{i}\dc{j}$, or $\L{i}\ec{j}$. Thus, mass suppression factors
for the first and second quark and lepton generations can appear
even at very low order as a result of the different weights of the
Higgs components.
In this model we found that, while the top quark can receive a 
viable, unsuppressed mass (given realistic Higgs VEVs),
the bottom quark mass, most second generation and some 
first generation masses were too small. This was, a result of the
weight factors for one $h$ component and for one $\bar{h}$ component, 
being too low. Phenomenology would have improved significantly if
the $\h{4}$ and $\hb{4}$ weights in $h$ and $\bar{h}$, respectively,
were larger by a factor of 100  
than their values of $10^{-5}$ and $10^{-4}$ found for our 
best non-Abelian flat direction, FDNA(5+8).
Also, we have observed the emergence of new techniques for the removal of
dangerous terms from $\vev{W}$ and from $\vev{F}$.  In Table 1B, four of our
flat directions are lifted to all order by the vanishing of terms with
more than two \NA fields.  Non-Abelian self-cancellation within single terms
is another promising tool for extending the order to which a direction is safe. 

The existence of Minimal Standard Heterotic String Models, which contain
solely the three generations of MSSM quarks and leptons and
a pair of Higgs doublets as the massless SM-charged states   
in the low energy effective field theory, has been a
significant discovery. Our FNY model has been the first example MSHSM.
Clearly, though the stringently flat $F$- and $D$-flat directions
that we have found can produce this MSHSM, they do not themselves lead to
viable quark and lepton mass matrices. Nevertheless, we have found
that these flat directions can present some interesting phenomenological
features such as multi-component physical Higgs that couple differently
to given quarks and leptons. One direction suggested by the 
partial success of these flat directions is investigation of non-stringently
flat directions for the FNY model that are flat to a finite order due to 
cancellation between various components in an $F$-term. The necessity of
non-stringent $F$-flatness was recently also shown for free fermionic
flipped $SU(5)$ models \cite{cen1}. 
Further, our discovery of an MSHSM in the 
neighborhood of the string/M-theory parameter space allowing  
free-fermionic description strongly suggests a search for 
further, perhaps more phenomenologically realistic, MSHSMs in this
region. This we leave for future research. 

\section{Acknowledgments}
This work is supported in part
by PPARC advanced fellowship (AF)
and DE-FG-0395ER40917 (GC,DVN,JW).
GC thanks John Rizos for very helpful discussion
regarding application of global worldsheet charge constraints 
for superpotential terms of high order. 
\newpage
\appendix

\section{Dimension-1 Non-Abelian $D$-flat MSSM Directions} 

\def\ify{$\infty$ }
\def\ifw{${\infty}^{\ast}$ }

\def\x{ $\phantom{1}$}
\def\y{$\ast$}
\def\my{$\bar{\ast}$}
\def\ny{${(-)}\atop{\ast}$}

\begin{flushleft}
\begin{tabular}{|l|r||r||r|r|r|r|r|}
\hline
\hline
FD$\#$&${\cal{O}}(W)$
      &Q'&$\p{23}(\p{4})\Hs{38,31}$&$\bp{56}\pp{56}\sH{30,21,15,17,36,37}$
                                                                      &$\Nc{i=1,2,3}$
                                                                      &$\H{28}\V{40}$
                                                                      &$\H{23,26}\V{5,7,35,37}$\\
\hline
 5&$\infty$
     &-1&  2\,   2\,  1\,  1& 0\,   0\,   4\,   0\,  3\, 0\, 0\, 0& 0\, 0\, 1 &0\,0 &  0\,1\, 0\,0\,1\, 0 \\
 6&$\infty$
     &-1&  3\,   3\,  1\,  1& 0\,   0\,   5\,   0\,  3\, 0\, 0\, 0& 1\, 1\, 0 &0\,0 &  0\,2\, 0\,0\,2\, 0 \\
 7&$\infty$
     &-2&  7\,  10\,  2\,  4& 1\,   0\,  12\,   0\,  8\, 0\, 0\, 0& 6\, 0\, 0 &0\,0 &  0\,6\, 0\,0\,0\, 6 \\
 8&$\infty$
     &-2&  9\,  22\,  2\,  4& 0\,  -1\,  16\,   8\,  0\, 0\,0\, 10& 0\, 0\, 0 &0\,0 &  0\,0\,18\,0\,0\,18 \\
 9&$\infty$
     &-2&  9\,   6\,  2\,  4& 0\,  -1\,  16\,   0\,  0\, 8\,0\, 10& 0\, 0\, 0 &0\,0 &  0\,0\,10\,0\,0\,10 \\
10&10&-2&  6\,   4\,  2\,  2& 0\,   0\,   8\,   0\,  4\, 0\, 0\, 0& 0\, 4\, 0 &0\,0 &  0\,4\, 0\,0\,2\, 2 \\
11& 9&-1&  4\,   2\,  1\,  1& 0\,  -1\,   4\,   0\,  0\, 0\, 0\, 0& 0\, 4\, 0 &0\,0 &  0\,4\, 0\,0\,1\, 3 \\
12& 9&-2&  8\,  -4\,  2\,  4& 1\,  -1\,   6\,   0\,  0\, 0\, 0\, 0& 0\, 8\, 0 &8\,8 &  0\,0\, 0\,0\,0\, 0 \\
13& 8&-2&  8\,   4\,  2\,  4& 1\,  -1\,   6\,   0\,  0\, 0\, 0\, 0& 0\, 8\, 0 &0\,0 &  0\,8\, 0\,0\,0\, 8 \\
14& 8&-1&  7\,   5\,  4\,  8& 5\,   1\,   0\,   0\,  0\, 0\, 0\, 0& 0\, 4\, 0 &0\,0 &  3\,7\, 0\,0\,0\,10 \\
15& 8&-1&  5\,   3\,  2\,  4& 2\,   0\,   2\,   0\,  0\, 0\, 0\, 0& 0\, 4\, 0 &0\,0 &  1\,5\, 0\,0\,0\, 6 \\
16& 8&-2&  7\,   4\,  2\,  4& 1\,   0\,   6\,   0\,  2\, 0\, 0\, 0& 0\, 6\, 0 &0\,0 &  0\,6\, 0\,0\,0\, 6 \\
17& 7&-4& 15\,   6\,  2\,  4& 0\,  -3\,  14\,   0\,  0\, 0\, 0\, 0&0\, 16\, 0 &0\,0 & 0\,16\, 0\,2\,0\,14 \\
18& 7&-2&  7\,   5\,  3\,  1& 0\,   0\,  10\,   0\,  6\, 0\, 1\, 0& 0\, 2\, 0 &0\,0 &  0\,4\, 0\,0\,4\, 0 \\
\hline
5+8
 & 5&-20& 45\,  58\, 20\, 22& 0\,  -1\,  88\,   8\, 54\, 0\,0\,10 & 0\,0\,18 &0\,0 & 0\,18\,18\,0\,18\,18 \\
\hline
\end{tabular}
\end{flushleft}
\no Table 1A: FNY directions flat through at least seventh order that
contain VEVs of Non-Abelian charged Hidden Sector Fields.
\no All component VEVs in these directions are uncharged under the MSSM gauge group.
Column one entries specify the class to which an example direction belongs.
Column two entries give the anomalous charges $Q'\equiv Q^{(A)}/112$ of the
flat directions.
The next several column entries specify the ratios of the norms of the VEV$^2$s.
The $\p{4}$-related component is the net value (in units of the square
overall VEV scale) of
$\mvev{\Phi_4} + \mvev{\Phi^{'}_4} - \mvev{\bar{\Phi}_4} - \mvev{\bar{\Phi}^{'}_4}$.
E.g., a ``1'' in the $\Phi_4$ column for FDNA1 specifies that
$\mvev{\Phi_4} + \mvev{\Phi^{'}_4} - \mvev{\bar{\Phi}_4} - \mvev{\bar{\Phi}^{'}_4} =
1\times \mvev{\alpha}$.
For consistency, we continue here the notation of \cite{cfn4}. However, since
FDNA1 through FDNA4 contain the unwanted VEV $\vev{\p{12}}$, they are excluded
and we begin with FDNA5.

\begin{flushleft}
\begin{tabular}{|l|r||l||l|}
\hline
\hline
FD$\#$& ${\cal O}(W)$ & Dangerous $W$ Terms & Comments\\
\hline
 5& $\infty$ & None & No gauge invariant $\vev{F}$ terms \\
 6& $\infty$ & None & ${13}^{th}$ order NAFE \\
 7& $\infty$ & None & ${13}^{th}$ order NAFE \\
 8& $\infty$ & None & Safe by string selection rules \\
 9& $\infty$ & None & Safe by string selection rules \\
10& $\infty$ & None & ${10}^{th}$ and $12^{th}$ order NAFE \\
11&  9       & $\vev{\p{23} (\p{4}) \sH{38} \bi{56} \sH{30} \Nc{2} \H{26}\cdot \V{35}}\sH{20}$ & No NASC \\
12&  9       & $\vev{\p{23} (\bp{4}) \sH{38} \sH{31} \bp{56} \Nc{2} \H{28}\cdot \V{40}}\sH{22}$ & No NASC \\
13,14,15,16
  &8         & $\vev{\p{23} \sH{38} \sH{31} \bp{56} \Nc{2} \H{26}\cdot \V{37}}\sH{22}$& No NASC \\
17& $\infty$ & None & $7^{th}$ and ${12}^{th}$ order NAFE \\
18&  7       & $\vev{\p{23} \sH{30} \sH{15} \H{36} \Nc{2} \H{26}\cdot }\V{7}$ & No NASC \\
\hline
5+8& 5       & $\vev{\pp{4} \sH{30} \sH{21} \V{35} \cdot }V_{15}$ & No NASC \\
   & 5       & $\vev{\pp{4} \sH{30} \sH{15} \V{5} \cdot }V_{17}$ & No NASC \\
\hline
\hline
\end{tabular}
\end{flushleft}

\no Table 1B: All superpotential terms which break $F$-flatness in the
Table 1A directions.
\no Column one entries specify the class of a flat direction.
The entry in the next column specifies superpotential terms which break
$F$-flatness.
For the final column, the notation ``NAFE'' indicates that the direction was
made safe by the removal of terms at the specified order(s) with a \NA field
excess (always four total for these cases).  ``No NASC'' means that
$D$-constraints
would not permit a consistent \NA self-cancellation solution.  For FDNA(5+8),
which breaks at order five, there are self-cancellations as low as order four.
For details, see Table \ref{danger8} and the surrounding discussion of Section
\ref{58FD}.

\section{Potential FNY Quark and Charged Lepton Mass Terms (through
8th order)}

\no Possible up--quark mass terms from $\vev{\hb{1}}$:
\beqn 
{
\begin{tabular}{lll}
   $\pps\bh{1}\, [\Q{1 }  \uc{1} $&&\\
   $+ \Q{1 }  \uc{2} \sH{15} \sH{30} \sV{11} \sV{12} $  
 & $+ \Q{1 }  \uc{2} \sH{15} \sH{30}  \V{14}  \V{13} $  
 & $+ \Q{1 }  \uc{2} \sH{15} \sH{30}  \V{19}  \V{20} $\\ 
   $+ \Q{1 }  \uc{2} \sH{15} \sH{30}  \V{15}  \V{17} $  
 & $+ \Q{1 }  \uc{3} \sH{15} \sH{30} \sV{11} \sV{22} $  
 & $+ \Q{1 }  \uc{3} \sH{15} \sH{30}  \V{14}  \V{23} $\\ 
   $+ \Q{1 }  \uc{3} \sH{30} \sH{38}  \V{15}  \V{35} $  
 & $+ \Q{3 }  \uc{1}  \Nc{2} \sH{30} \sH{36} \sV{31} $  
 & $+ \Q{3 }  \uc{1} \sH{17} \sH{30}  \V{19}  \V{30} $\\ 
   $+ \Q{3 }  \uc{1} \sH{17} \sH{30}  \V{15}  \V{27} $  
 & $+ \Q{3 }  \uc{1} \sH{30} \sH{38}  \V{19}  \V{39} $  
 & $+ \Q{3 }  \uc{3}  \V{30}  \V{40}  \V{27}  \V{37} $\\ 
   $+ \Q{1 }  \uc{2}  \p{23} \sH{15} \sH{30} \sH{36} \sH{37} $  
 & $+ \Q{1 }  \uc{2}  \p{23} \sH{15} \sH{30} \sV{21} \sV{22} $  
 & $+ \Q{1 }  \uc{2}  \p{23} \sH{15} \sH{30}  \V{24}  \V{23} $\\ 
   $+ \Q{1 }  \uc{2}  \p{23} \sH{15} \sH{30}  \V{29}  \V{30} $  
 & $+ \Q{1 }  \uc{2}  \p{23} \sH{15} \sH{30}  \V{25}  \V{27} $  
 & $+ \Q{1 }  \uc{2}  \p{23} \sH{30} \sH{37}  \V{25}  \V{37} $\\ 
   $+ \Q{1 }  \uc{2} \bp{56} \sH{32} \sH{38} \sV{21} \sV{32} $  
 & $+ \Q{1 }  \uc{2} \bp{56} \sH{38} \sV{32}  \V{24}  \H{35} $  
 & $+ \Q{1 }  \uc{3}  \Nc{2} \bp{13} \sH{30}  \H{26}  \V{37} $\\
\end{tabular}\nolabel}
\eeqn
\newpage

\no Possible up--quark mass terms from $\vev{\bh{2}}$:
\beqn 
{\begin{tabular}{lll}
   $\pps\bh{2}\, [\Q{1 }  \uc{1} \sH{29} \sH{30} \sV{31} \sV{32} $  
 & $+ \Q{1 }  \uc{2} \sH{15} \sH{30} \sV{31} \sV{32} $  
 & $+ \Q{3 }  \uc{3} \sH{19} \sH{30}  \V{15}  \V{37} $\\
   $+ \Q{1 }  \uc{1}  \Nc{1}  \p{12} \sH{30} \sH{32} \sV{1 } $  
 & $+ \Q{1 }  \uc{1}  \Nc{3}  \p{12} \sH{16}  \H{26}  \V{37} $  
 & $+ \Q{1 }  \uc{1}  \Nc{3}  \p{12} \sH{30} \sH{32} \sV{21} $\\ 
   $+ \Q{1 }  \uc{1}  \Nc{3}  \p{12} \sH{30}  \V{24}  \H{35} $  
 & $+ \Q{1 }  \uc{1}  \p{12} \sH{15} \sH{30}  \V{14}  \V{3 } $  
 & $+ \Q{1 }  \uc{1}  \p{12} \sH{17} \sH{30}  \V{4 }  \V{13} $\\ 
   $+ \Q{1 }  \uc{1}  \p{12} \sH{17} \sH{30}  \V{9 }  \V{19} $  
 & $+ \Q{1 }  \uc{1}  \p{12} \sH{19} \sH{31}  \V{15}  \V{37} $  
 & $+ \Q{1 }  \uc{1}  \p{12} \sH{19} \sH{32}  \V{34}  \V{13} $\\ 
   $+ \Q{1 }  \uc{1}  \p{12} \sH{19} \sV{12}  \V{34}  \H{35} $  
 & $+ \Q{1 }  \uc{1}  \p{12} \sH{21} \sH{30}  \V{19}  \V{39} $  
 & $+ \Q{1 }  \uc{1}  \p{12} \sH{30} \sH{32} \sH{37} \sH{39} $\\ 
   $+ \Q{1 }  \uc{1}  \p{12} \sH{30} \sH{37}  \H{42}  \H{35} $  
 & $+ \Q{1 }  \uc{1}  \p{12} \sH{30} \sV{31}  \H{23}  \V{35} $  
 & $+ \Q{1 }  \uc{1}  \p{12} \sH{37} \sV{32}  \H{28}  \V{40} $\\ 
   $+ \Q{1 }  \uc{3}  \Nc{2}  \p{23} \sH{30}  \H{26}  \V{37} $  
 & $+ \Q{1 }  \uc{3}  \p{12} \sH{15} \sH{30} \sV{11} \sV{22} $  
 & $+ \Q{1 }  \uc{3}  \p{12} \sH{15} \sH{30}  \V{14}  \V{23} $\\ 
   $+ \Q{1 }  \uc{3}  \p{12} \sH{30} \sH{38}  \V{15}  \V{35} $  
 & $+ \Q{1 }  \uc{3}  \p{12}  \V{9 }  \V{40}  \V{27}  \V{37} $  
 & $+ \Q{1 }  \uc{3} \bp{56} \sH{21} \sH{30} \sH{32} \sH{39} $\\
\end{tabular}\nolabel}
\eeqn
\newpage

\no Possible up--quark mass terms from $\vev{\bh{3}}$:
\beqn 
{\begin{tabular}{lll}
   $\pps\bh{3}\, [ \Q{1 }  \uc{1} \sH{29} \sH{30} $  
 & $+ \Q{1 }  \uc{2} \sH{15} \sH{30} $&\\ 
   $+ \Q{2 }  \uc{1}  \pp{4} \sH{15} \sH{30} $&&\\ 
   $+ \Q{1 }  \uc{3}  \Nc{2} \sH{30}  \H{26}  \V{37} $  
 & $+ \Q{3 }  \uc{1}  \Nc{2} \sH{30}  \H{28}  \V{40} $\\  
   $+ \Q{1 }  \uc{1}  \Nc{1}  \p{13} \sH{30} \sH{32} \sV{1 } $  
 & $+ \Q{1 }  \uc{1}  \Nc{3}  \p{13} \sH{16}  \H{26}  \V{37} $  
 & $+ \Q{1 }  \uc{1}  \Nc{3}  \p{13} \sH{30} \sH{32} \sV{21} $\\ 
   $+ \Q{1 }  \uc{1}  \Nc{3}  \p{13} \sH{30}  \V{24}  \H{35} $  
 & $+ \Q{1 }  \uc{1}  \p{13} \sH{15} \sH{30}  \V{14}  \V{3 } $  
 & $+ \Q{1 }  \uc{1}  \p{13} \sH{17} \sH{30}  \V{4 }  \V{13} $\\ 
   $+ \Q{1 }  \uc{1}  \p{13} \sH{17} \sH{30}  \V{9 }  \V{19} $  
 & $+ \Q{1 }  \uc{1}  \p{13} \sH{19} \sH{31}  \V{15}  \V{37} $  
 & $+ \Q{1 }  \uc{1}  \p{13} \sH{19} \sH{32}  \V{34}  \V{13} $\\ 
   $+ \Q{1 }  \uc{1}  \p{13} \sH{19} \sV{12}  \V{34}  \H{35} $  
 & $+ \Q{1 }  \uc{1}  \p{13} \sH{21} \sH{30}  \V{19}  \V{39} $  
 & $+ \Q{1 }  \uc{1}  \p{13} \sH{30} \sH{32} \sH{37} \sH{39} $\\ 
   $+ \Q{1 }  \uc{1}  \p{13} \sH{30} \sH{37}  \H{42}  \H{35} $  
 & $+ \Q{1 }  \uc{1}  \p{13} \sH{30} \sV{31}  \H{23}  \V{35} $  
 & $+ \Q{1 }  \uc{1}  \p{13} \sH{37} \sV{32}  \H{28}  \V{40} $\\
\end{tabular}\nolabel}
\eeqn
\newpage

\no Possible up--quark mass terms from $\vev{\bh{4}\equiv\H{34}}$:
\beqn 
{\begin{tabular}{lll}
   $\pps\h{4}\, [ \Q{3 }  \uc{3} \sH{30} $&&\\ 
   $+ \Q{1 }  \uc{1}  \p{12} \sH{31} $  
 & $+ \Q{2 }  \uc{1}  \p{12} \sH{16} $&\\ 
   $+ \Q{1 }  \uc{2}  \p{12} \bpp{4} \sH{16} $\\ 
   $+ \Q{1 }  \uc{1}  \p{13} \sH{31} \sV{31} \sV{32} $  
 & $+ \Q{1 }  \uc{1}  \p{23} \sH{29} \sH{30} \sH{31} $  
 & $+ \Q{1 }  \uc{2}  \p{23} \sH{15} \sH{30} \sH{31} $\\
   $+ \Q{1 }  \uc{3}  \p{12} \sH{30} \sV{2 } \sV{21} $  
 & $+ \Q{1 }  \uc{3}  \p{12} \sH{30}  \V{9 }  \V{29} $  
 & $+ \Q{2 }  \uc{1}  \p{13} \sH{16} \sV{31} \sV{32} $\\ 
   $+ \Q{2 }  \uc{1}  \p{23} \sH{16} \sH{29} \sH{30} $  
 & $+ \Q{2 }  \uc{2}  \p{23} \sH{15} \sH{16} \sH{30} $  
 & $+ \Q{2 }  \uc{2} \bi{56} \sH{30} \sV{31} \sV{32} $\\ 
   $+ \Q{2 }  \uc{3}  \p{12} \sH{30} \sV{12} \sV{21} $  
 & $+ \Q{2 }  \uc{3}  \p{12} \sH{30}  \V{24}  \V{13} $  
 & $+ \Q{3 }  \uc{1}  \p{12} \sH{22} \sH{30} \sH{37} $\\ 
   $+ \Q{3 }  \uc{1}  \p{12} \sH{30}  \V{24}  \V{3 } $  
 & $+ \Q{3 }  \uc{1}  \p{12} \sH{30}  \V{7 }  \V{25} $  
 & $+ \Q{3 }  \uc{2}  \p{12} \sH{30} \sV{12} \sV{21} $\\
\end{tabular}\nolabel}
\eeqn
\newpage

\no Possible down--quark mass terms from $\vev{\h{1}}$:
\beqn 
{\begin{tabular}{lll}
   $\pps\h{1}\, [ \Q{3 }  \dc{3} \sH{29} \sH{30} $&&\\ 
   $+ \Q{2 }  \dc{2} \sH{29} \sH{30} \sV{31} \sV{32} $&&\\  
   $+ \Q{1 }  \dc{2}  \Nc{3}  \p{12} \sH{31}  \H{26}  \V{37} $  
 & $+ \Q{1 }  \dc{2}  \p{12} \sH{29} \sH{30} \sV{2 } \sV{11} $  
 & $+ \Q{1 }  \dc{2}  \p{12} \sH{29} \sH{30}  \V{5 }  \V{17} $\\ 
   $+ \Q{1 }  \dc{2}  \p{12} \sH{29} \sV{32}  \H{28}  \V{30} $  
 & $+ \Q{1 }  \dc{2}  \p{12} \sH{29} \sV{32}  \H{26}  \V{27} $  
 & $+ \Q{1 }  \dc{3}  \Nc{2}  \p{12} \sH{31}  \H{26}  \V{37} $\\ 
   $+ \Q{1 }  \dc{3}  \p{12} \sH{29} \sH{30} \sV{2 } \sV{21} $  
 & $+ \Q{1 }  \dc{3}  \p{12} \sH{29} \sH{30}  \V{9 }  \V{29} $  
 & $+ \Q{2 }  \dc{1}  \Nc{2}  \p{12} \sH{30} \sH{32} \sV{1 } $\\ 
   $+ \Q{2 }  \dc{1}  \p{12} \sH{29} \sH{30} \sV{1 } \sV{12} $  
 & $+ \Q{2 }  \dc{1}  \p{12} \sH{29} \sH{30}  \V{7 }  \V{15} $  
 & $+ \Q{2 }  \dc{2}  \Nc{1}  \p{12} \sH{30} \sH{32} \sV{1 } $\\ 
   $+ \Q{2 }  \dc{2}  \Nc{3}  \p{12} \sH{16}  \H{26}  \V{37} $  
 & $+ \Q{2 }  \dc{2}  \Nc{3}  \p{12} \sH{30} \sH{32} \sV{21} $  
 & $+ \Q{2 }  \dc{2}  \Nc{3}  \p{12} \sH{30}  \V{24}  \H{35} $\\ 
   $+ \Q{2 }  \dc{2}  \p{12} \sH{15} \sH{30}  \V{14}  \V{3 } $  
 & $+ \Q{2 }  \dc{2}  \p{12} \sH{17} \sH{30}  \V{4 }  \V{13} $  
 & $+ \Q{2 }  \dc{2}  \p{12} \sH{17} \sH{30}  \V{9 }  \V{19} $\\
\end{tabular}\nolabel}
\eeqn
\newpage


\no Possible electron mass terms from $\vev{\h{1}}$:
\beqn 
{\begin{tabular}{lll}
   $\pps\h{1}\, [ \L{3 }  \ec{3} \sH{29} \sH{30} $&&\\ 
   $+ \L{2 }  \ec{2} \sH{29} \sH{30} \sV{31} \sV{32} $\\  
   $+ \L{1 }  \ec{2}  \Nc{2}  \p{12} \sH{30}  \V{4 }  \H{35} $  
 & $+ \L{1 }  \ec{2}  \p{12} \sH{29} \sH{30}  \V{4 }  \V{13} $  
 & $+ \L{1 }  \ec{2}  \p{12} \sH{29} \sH{30}  \V{9 }  \V{19} $\\ 
   $+ \L{1 }  \ec{3}  \Nc{2}  \p{12} \sH{30}  \H{26}  \V{35} $  
 & $+ \L{1 }  \ec{3}  \Nc{3}  \p{13} \sH{30}  \V{4 }  \H{35} $  
 & $+ \L{1 }  \ec{3}  \p{12} \sH{20} \sH{29} \sH{30} \sH{37} $\\ 
   $+ \L{1 }  \ec{3}  \p{12} \sH{29} \sH{30}  \V{4 }  \V{23} $  
 & $+ \L{1 }  \ec{3}  \p{12} \sH{29} \sH{30}  \V{5 }  \V{27} $  
 & $+ \L{2 }  \ec{1}  \Nc{3}  \p{12} \sH{31}  \H{28}  \V{40} $\\ 
   $+ \L{2 }  \ec{1}  \p{12} \sH{29} \sH{30}  \V{14}  \V{3 } $  
 & $+ \L{2 }  \ec{2}  \Nc{1}  \p{12} \sH{30} \sH{32} \sV{1 } $  
 & $+ \L{2 }  \ec{2}  \Nc{3}  \p{12} \sH{16}  \H{26}  \V{37} $\\ 
   $+ \L{2 }  \ec{2}  \Nc{3}  \p{12} \sH{30} \sH{32} \sV{21} $  
 & $+ \L{2 }  \ec{2}  \Nc{3}  \p{12} \sH{30}  \V{24}  \H{35} $  
 & $+ \L{2 }  \ec{2}  \p{12} \sH{15} \sH{30}  \V{14}  \V{3 } $\\ 
   $+ \L{2 }  \ec{2}  \p{12} \sH{17} \sH{30}  \V{4 }  \V{13} $  
 & $+ \L{2 }  \ec{2}  \p{12} \sH{17} \sH{30}  \V{9 }  \V{19} $  
 & $+ \L{2 }  \ec{2}  \p{12} \sH{19} \sH{31}  \V{15}  \V{37} $\\
\end{tabular}\nolabel}
\eeqn
\newpage

 
\no Possible down--quark mass terms from $\vev{\h{2}}$:
\beqn 
{\begin{tabular}{lll}
   $\pps\h{2}\, [ \Q{2 }  \dc{2} $&&\\ 
   $+ \Q{3 }  \dc{3}  \V{34}  \V{33} $&&\\ 
   $+ \Q{1 }  \dc{3}  \p{13} \sV{2 } \sV{21} $ 
 & $+ \Q{1 }  \dc{3}  \p{13}  \V{9 }  \V{29} $  
 & $+ \Q{3 }  \dc{1}  \p{13} \sV{1 } \sV{22} $\\ 
   $+ \Q{1 }  \dc{1} \sH{29} \sH{30} \sV{1 } \sV{2 } $  
 & $+ \Q{1 }  \dc{1} \sH{29} \sH{30}  \V{4 }  \V{3 } $  
 & $+ \Q{1 }  \dc{1} \sH{29} \sH{30}  \V{5 }  \V{7 } $\\ 
   $+ \Q{1 }  \dc{2}  \Nc{3} \sH{31}  \H{26}  \V{37} $  
 & $+ \Q{1 }  \dc{2} \sH{15} \sH{19} \sH{20} \sH{31} $  
 & $+ \Q{1 }  \dc{2} \sH{19} \sH{31}  \V{5 }  \V{37} $\\ 
   $+ \Q{1 }  \dc{2} \sH{19} \sV{2 }  \V{34}  \H{35} $  
 & $+ \Q{1 }  \dc{2} \sH{29} \sV{32}  \H{28}  \V{30} $  
 & $+ \Q{1 }  \dc{2} \sH{29} \sV{32}  \H{26}  \V{27} $\\ 
   $+ \Q{1 }  \dc{3}  \Nc{2} \sH{31}  \H{26}  \V{37} $  
 & $+ \Q{1 }  \dc{3}  \p{13} \bp{4 } \sH{22} \sH{37} $  
 & $+ \Q{1 }  \dc{3}  \p{13} \bp{4 }  \V{24}  \V{3 } $\\
   $+ \Q{1 }  \dc{3}  \p{13} \bp{4 }  \V{7 }  \V{25} $  
 & $+ \Q{1 }  \dc{3} \sH{29} \sH{30} \sV{2 } \sV{21} $  
 & $+ \Q{1 }  \dc{3} \sH{29} \sH{30}  \V{9 }  \V{29} $\\
\end{tabular}\nolabel}
\eeqn
\newpage

\no Possible down--quark mass terms from $\vev{\h{2}}$ continued:
\beqn 
{\begin{tabular}{lll}

   $+ \Q{2 }  \dc{3} \bi{56} \sH{17} \sH{30}  \V{4 }  \V{23} $  
 & $+ \Q{2 }  \dc{3} \bi{56} \sH{17} \sH{30}  \V{5 }  \V{27} $  
 & $+ \Q{2 }  \dc{3} \bi{56} \sH{19} \sH{30}  \V{25}  \V{35} $\\ 
   $+ \Q{2 }  \dc{3} \bi{56} \sH{19} \sH{32}  \V{34}  \V{23} $  
 & $+ \Q{2 }  \dc{3} \bi{56} \sH{19} \sV{22}  \V{34}  \H{35} $  
 & $+ \Q{2 }  \dc{3} \bi{56} \sH{30} \sH{39}  \V{4 }  \V{33} $\\ 
   $+ \Q{3 }  \dc{1}  \Nc{2}  \pp{4} \sH{30}  \H{26}  \V{35} $  
 & $+ \Q{3 }  \dc{1}  \Nc{3} \bp{23} \sH{30} \sH{32} \sV{1 } $  
 & $+ \Q{3 }  \dc{1}  \p{12} \sV{1 } \sV{22}  \V{34}  \V{33} $\\ 
   $+ \Q{3 }  \dc{1}  \p{13} \sV{22} \sV{31}  \V{34}  \V{3 } $  
 & $+ \Q{3 }  \dc{1}  \p{13}  \V{30}  \V{39}  \V{7 }  \V{37} $  
 & $+ \Q{3 }  \dc{1}  \p{13}  \V{30}  \V{40}  \V{7 }  \V{35} $\\ 
   $+ \Q{3 }  \dc{1}  \p{4 } \sH{20} \sH{29} \sH{30} \sH{37} $  
 & $+ \Q{3 }  \dc{1}  \p{4 } \sH{29} \sH{30}  \V{4 }  \V{23} $  
 & $+ \Q{3 }  \dc{1}  \p{4 } \sH{29} \sH{30}  \V{5 }  \V{27} $\\ 
   $+ \Q{3 }  \dc{1} \bp{56} \sH{19} \sH{29} \sH{31} \sH{36} $  
 & $+ \Q{3 }  \dc{2}  \Nc{1}  \pp{4} \sH{30}  \H{26}  \V{35} $  
 & $+ \Q{3 }  \dc{2}  \Nc{2} \pp{56} \sH{16}  \H{26}  \V{37} $\\ 
   $+ \Q{3 }  \dc{2}  \p{12} \sV{11} \sV{22}  \V{34}  \V{33} $  
 & $+ \Q{3 }  \dc{2}  \p{12}  \V{14}  \V{34}  \V{23}  \V{33} $  
 & $+ \Q{3 }  \dc{2}  \p{13} \sH{19} \sH{20} \sV{12} \sV{21} $\\ 
   $+ \Q{3 }  \dc{2}  \p{13} \sH{19} \sH{20}  \V{24}  \V{13} $  
 & $+ \Q{3 }  \dc{2}  \p{13} \sH{19} \sH{36}  \V{4 }  \V{13} $  
 & $+ \Q{3 }  \dc{2}  \p{13} \sH{19} \sH{36}  \V{9 }  \V{19} $\\
\end{tabular}\nolabel}
\eeqn
\newpage


\no Possible electron mass terms from $\vev{\h{2}}$:
\beqn 
{\begin{tabular}{lll}
   $\pps\h{2}\, [  \L{2 }  \ec{2} $&&\\ 
   $+ \L{3 }  \ec{3}  \V{34}  \V{33} $&&\\
   $+ \L{1 }  \ec{3}  \p{13} \sH{20} \sH{37} $ 
 & $+ \L{1 }  \ec{3}  \p{13}  \V{4 }  \V{23} $  
 & $+ \L{1 }  \ec{3}  \p{13}  \V{5 }  \V{27} $\\ 
   $+ \L{3 }  \ec{1}  \p{13} \sH{22} \sH{37} $ 
 & $+ \L{3 }  \ec{1}  \p{13}  \V{24}  \V{3 } $  
 & $+ \L{3 }  \ec{1}  \p{13}  \V{7 }  \V{25} $\\ 
   $+ \L{1 }  \ec{1} \sH{29} \sH{30} \sV{1 } \sV{2 } $  
 & $+ \L{1 }  \ec{1} \sH{29} \sH{30}  \V{4 }  \V{3 } $  
 & $+ \L{1 }  \ec{1} \sH{29} \sH{30}  \V{5 }  \V{7 } $\\ 
   $+ \L{1 }  \ec{2} \sH{15} \sH{30} \sV{1 } \sV{2 } $  
 & $+ \L{1 }  \ec{2} \sH{15} \sH{30}  \V{4 }  \V{3 } $  
 & $+ \L{1 }  \ec{2} \sH{15} \sH{30}  \V{5 }  \V{7 } $\\ 
   $+ \L{1 }  \ec{3}  \Nc{2} \sH{30}  \H{26}  \V{35} $  
 & $+ \L{1 }  \ec{3}  \p{13} \bp{4 } \sV{1 } \sV{22} $  
 & $+ \L{1 }  \ec{3} \sH{20} \sH{29} \sH{30} \sH{37} $\\ 
   $+ \L{1 }  \ec{3} \sH{29} \sH{30}  \V{4 }  \V{23} $  
 & $+ \L{1 }  \ec{3} \sH{29} \sH{30}  \V{5 }  \V{27} $  
 & $+ \L{2 }  \ec{1}  \Nc{3} \sH{31}  \H{28}  \V{40} $\\
\end{tabular}\nolabel}
\eeqn
\newpage

\no Possible electron mass terms from $\vev{\h{2}}$ continued:
\beqn 
{\begin{tabular}{lll}

   $+ \L{2 }  \ec{3}  \p{4 } \sH{30}  \H{42}  \V{4 }  \V{34} $  
 & $+ \L{2 }  \ec{3} \bp{4 } \sH{17} \sH{30} \sV{1 } \sV{22} $  
 & $+ \L{2 }  \ec{3} \bp{4 } \sH{19} \sH{30}  \V{29}  \V{39} $\\ 
   $+ \L{2 }  \ec{3} \bp{4 } \sH{30} \sV{1 }  \H{42}  \V{33} $  
 & $+ \L{2 }  \ec{3} \bpp{4} \sH{19}  \H{35}  \V{23}  \V{33} $  
 & $+ \L{2 }  \ec{3} \bp{56} \sH{17} \sH{20} \sH{31} \sH{38} $\\ 
   $+ \L{2 }  \ec{3} \bp{56} \sH{18} \sH{19} \sH{29} \sH{36} $  
 & $+ \L{2 }  \ec{3} \pp{56} \sH{15} \sH{22} \sH{30} \sH{37} $  
 & $+ \L{2 }  \ec{3} \pp{56} \sH{15} \sH{30}  \V{24}  \V{3 } $\\ 
   $+ \L{2 }  \ec{3} \pp{56} \sH{15} \sH{30}  \V{7 }  \V{25} $  
 & $+ \L{3 }  \ec{1}  \Nc{1}  \p{13} \sH{16}  \H{28}  \V{39} $  
 & $+ \L{3 }  \ec{1}  \Nc{2}  \p{13} \sH{31} \sH{36} \sV{31} $\\ 
   $+ \L{3 }  \ec{1}  \Nc{2}  \p{4 } \sH{31}  \H{26}  \V{37} $  
 & $+ \L{3 }  \ec{1}  \p{12} \sH{22} \sH{37}  \V{34}  \V{33} $  
 & $+ \L{3 }  \ec{1}  \p{12}  \V{24}  \V{34}  \V{3 }  \V{33} $\\ 
   $+ \L{3 }  \ec{1}  \p{12}  \V{34}  \V{33}  \V{7 }  \V{25} $  
 & $+ \L{3 }  \ec{1}  \p{13} \sH{16} \sH{17} \sH{20} \sH{37} $  
 & $+ \L{3 }  \ec{1}  \p{13} \sH{16} \sH{17}  \V{4 }  \V{23} $\\ 
   $+ \L{3 }  \ec{1}  \p{13} \sH{16} \sH{17}  \V{5 }  \V{27} $  
 & $+ \L{3 }  \ec{1}  \p{13} \sH{16} \sH{19}  \V{25}  \V{35} $  
 & $+ \L{3 }  \ec{1}  \p{13} \sH{16} \sH{29} \sV{11} \sV{22} $\\ 
   $+ \L{3 }  \ec{1}  \p{13} \sH{16} \sH{29}  \V{14}  \V{23} $  
 & $+ \L{3 }  \ec{1}  \p{13} \sH{16} \sH{39}  \V{4 }  \V{33} $  
 & $+ \L{3 }  \ec{1}  \p{13} \sH{17} \sH{31}  \V{19}  \V{30} $\\
\end{tabular}\nolabel}
\eeqn
\newpage


\no Possible down--quark mass terms from $\vev{\h{3}}$:
\beqn 
{\begin{tabular}{lll}
   $\pps\h{3}\, [ \Q{3 }  \dc{3} $&&\\ 
   $+ \Q{2 }  \dc{2} \sV{31} \sV{32} $&&\\ 
   $+ \Q{1 }  \dc{2}  \p{12} \sV{2 } \sV{11} $  
 & $+ \Q{1 }  \dc{2}  \p{12}  \V{5 }  \V{17} $  
 & $+ \Q{2 }  \dc{1}  \p{12} \sV{1 } \sV{12} $\\ 
   $+ \Q{2 }  \dc{1}  \p{12}  \V{7}   \V{15} $\\ 
   $+ \Q{1 }  \dc{2}  \p{12} \bpp{4}  \V{14}  \V{3 } $  
 & $+ \Q{1 }  \dc{2} \sH{19} \sV{32}  \V{4 }  \H{35} $  
 & $+ \Q{1 }  \dc{3} \sH{15} \sV{32}  \H{26}  \V{7 } $\\ 
   $+ \Q{2 }  \dc{1}  \p{12}  \pp{4}  \V{4 }  \V{13} $  
 & $+ \Q{2 }  \dc{1}  \p{12}  \pp{4}  \V{9 }  \V{19} $  
 & $+ \Q{3 }  \dc{1}  \Nc{3} \sH{30} \sH{32} \sV{1 } $\\ 
   $+ \Q{3 }  \dc{2}  \Nc{3} \sH{30} \sH{32} \sV{11} $  
 & $+ \Q{3 }  \dc{2}  \Nc{3} \sH{30}  \V{14}  \H{35} $  
 & $+ \Q{3 }  \dc{2} \sH{30} \sH{39} \sV{2 } \sV{31} $\\ 
   $+ \Q{1 }  \dc{1}  \Nc{2}  \p{12} \sH{20} \sH{31} \sV{31} $  
 & $+ \Q{1 }  \dc{1}  \Nc{3}  \p{12} \sH{16}  \H{26}  \V{35} $  
 & $+ \Q{1 }  \dc{1}  \p{12} \sH{19} \sH{31}  \V{15}  \V{35} $\\
\end{tabular}\nolabel}
\eeqn
\newpage

\no Possible down--quark mass terms from $\vev{\h{3}}$ continued:
\beqn 
{\begin{tabular}{lll}
   $+ \Q{3 }  \dc{2}  \Nc{3} \bpp{4} \sH{22} \sH{30} \sV{31} $  
 & $+ \Q{3 }  \dc{2}  \p{12} \sH{19} \sH{20} \sV{12} \sV{21} $  
 & $+ \Q{3 }  \dc{2}  \p{12} \sH{19} \sH{20}  \V{24}  \V{13} $\\ 
   $+ \Q{3 }  \dc{2}  \p{12} \sH{19} \sH{36}  \V{4 }  \V{13} $  
 & $+ \Q{3 }  \dc{2}  \p{12} \sH{19} \sH{36}  \V{9 }  \V{19} $  
 & $+ \Q{3 }  \dc{2}  \p{12} \sH{20} \sH{38} \sV{1 } \sV{12} $\\ 
   $+ \Q{3 }  \dc{2}  \p{12} \sH{20} \sH{38}  \V{7 }  \V{15} $  
 & $+ \Q{3 }  \dc{2}  \p{12} \sH{20} \sV{31}  \H{25}  \V{30} $  
 & $+ \Q{3 }  \dc{2}  \p{12} \sH{20} \sV{31}  \H{23}  \V{27} $\\ 
   $+ \Q{3 }  \dc{2}  \p{12}  \V{20}  \V{39}  \V{27}  \V{37} $  
 & $+ \Q{3 }  \dc{2}  \p{12}  \V{20}  \V{40}  \V{27}  \V{35} $  
 & $+ \Q{3 }  \dc{2}  \p{12}  \V{30}  \V{39}  \V{17}  \V{37} $\\ 
   $+ \Q{3 }  \dc{2}  \p{12}  \V{30}  \V{40}  \V{17}  \V{35} $  
 & $+ \Q{3 }  \dc{2}  \p{13} \sV{11} \sV{22} \sV{31} \sV{32} $  
 & $+ \Q{3 }  \dc{2}  \p{13} \sV{31} \sV{32}  \V{14}  \V{23} $\\ 
   $+ \Q{3 }  \dc{2}  \p{23} \sH{17} \sH{20} \sH{30} \sH{37} $  
 & $+ \Q{3 }  \dc{2}  \p{23} \sH{17} \sH{30}  \V{4 }  \V{23} $  
 & $+ \Q{3 }  \dc{2}  \p{23} \sH{17} \sH{30}  \V{5 }  \V{27} $\\ 
   $+ \Q{3 }  \dc{2}  \p{23} \sH{19} \sH{30}  \V{25}  \V{35} $  
 & $+ \Q{3 }  \dc{2}  \p{23} \sH{19} \sH{32}  \V{34}  \V{23} $  
 & $+ \Q{3 }  \dc{2}  \p{23} \sH{19} \sV{22}  \V{34}  \H{35} $\\ 
   $+ \Q{3 }  \dc{2}  \p{23} \sH{29} \sH{30} \sV{11} \sV{22} $  
 & $+ \Q{3 }  \dc{2}  \p{23} \sH{29} \sH{30}  \V{14}  \V{23} $  
 & $+ \Q{3 }  \dc{2}  \p{23} \sH{30} \sH{39}  \V{4 }  \V{33} $\\ 
   $+ \Q{3 }  \dc{2} \bpp{4} \sH{30} \sV{31}  \H{42}  \V{3 } $  
 & $+ \Q{3 }  \dc{2} \bp{56} \sH{17} \sV{32}  \H{28}  \V{20} $  
 & $+ \Q{3 }  \dc{2} \bp{56} \sH{17} \sV{32}  \H{26}  \V{17} $\\ 
   $+ \Q{3 }  \dc{2} \bp{56} \sH{18} \sH{19} \sH{32} \sH{39} $  
 & $+ \Q{3 }  \dc{2} \bp{56} \sH{18} \sH{19}  \H{42}  \H{35} $  
 & $+ \dots]$\\ 
\end{tabular}\nolabel}
\eeqn
\newpage


\no Possible electron mass terms from $\vev{\h{3}}$:
\beqn 
{\begin{tabular}{lll}
   $\pps\h{3}\, [ \L{3 }  \ec{3} $&&\\ 
   $+ \L{2 }  \ec{2} \sV{31} \sV{32} $&&\\ 
   $+ \L{1 }  \ec{2}  \p{12}  \V{4 }  \V{13} $  
 & $+ \L{1 }  \ec{2}  \p{12}  \V{9 }  \V{19} $  
 & $+ \L{2 }  \ec{1}  \p{12}  \V{14}  \V{3 } $\\ 
   $+ \L{1 }  \ec{2}  \p{12} \bpp{4} \sV{1 } \sV{12} $  
 & $+ \L{1 }  \ec{2}  \p{12} \bpp{4}  \V{7 }  \V{15} $  
 & $+ \L{1 }  \ec{3}  \Nc{3} \sH{30}  \V{4 }  \H{35} $\\ 
   $+ \L{2 }  \ec{1}  \p{12}  \pp{4} \sV{2 } \sV{11} $  
 & $+ \L{2 }  \ec{1}  \p{12}  \pp{4}  \V{5 }  \V{17} $  
 & $+ \L{2 }  \ec{1} \sH{19} \sH{32} \sV{1 } \sV{32} $\\ 
   $+ \L{2 }  \ec{3}  \Nc{3} \sH{30} \sH{32} \sV{11} $  
 & $+ \L{2 }  \ec{3}  \Nc{3} \sH{30}  \V{14}  \H{35} $  
 & $+ \L{2 }  \ec{3} \sH{30} \sH{39} \sV{2 } \sV{31} $\\ 
   $+ \L{1 }  \ec{1}  \Nc{2}  \p{12} \sH{20} \sH{31} \sV{31} $  
 & $+ \L{1 }  \ec{1}  \Nc{3}  \p{12} \sH{16}  \H{26}  \V{35} $  
 & $+ \L{1 }  \ec{1}  \p{12} \sH{19} \sH{31}  \V{15}  \V{35} $\\ 
   $+ \L{1 }  \ec{1}  \p{12} \sH{37} \sV{32}  \H{28}  \V{39} $  
 & $+ \L{1 }  \ec{2}  \Nc{2}  \p{23} \sH{30}  \V{4 }  \H{35} $  
 & $+ \L{1 }  \ec{2}  \p{12} \sV{2 } \sV{31}  \V{34}  \V{13} $\\
\end{tabular}\nolabel}
\eeqn
\newpage

\no Possible electron mass terms from $\vev{\h{3}}$ continued:
\beqn 
{\begin{tabular}{lll}
   $+ \L{3 }  \ec{1}  \p{13} \sH{31} \sH{39} \sV{12} \sV{31} $  
 & $+ \L{3 }  \ec{1}  \p{13} \sH{31} \sV{31}  \H{42}  \V{13} $  
 & $+ \L{3 }  \ec{1}  \p{13} \sV{31} \sV{32}  \V{24}  \V{3 } $\\ 
   $+ \L{3 }  \ec{1}  \p{13} \sV{31} \sV{32}  \V{7 }  \V{25} $  
 & $+ \L{3 }  \ec{1}  \p{23} \sH{22} \sH{29} \sH{30} \sH{37} $  
 & $+ \L{3 }  \ec{1}  \p{23} \sH{29} \sH{30}  \V{24}  \V{3 } $\\ 
   $+ \L{3 }  \ec{1}  \p{23} \sH{29} \sH{30}  \V{7 }  \V{25} $  
 & $+ \L{3 }  \ec{1}  \p{4 } \sH{15} \sV{32}  \H{26}  \V{7 } $  
 & $+ \L{3 }  \ec{2}  \Nc{2}  \p{23} \sH{16}  \H{26}  \V{37} $\\ 
   $+ \L{3 }  \ec{2}  \Nc{2}  \p{23} \sH{30} \sH{32} \sV{21} $  
 & $+ \L{3 }  \ec{2}  \Nc{2}  \p{23} \sH{30}  \V{24}  \H{35} $  
 & $+ \L{3 }  \ec{2}  \Nc{3} \bi{56} \sH{30} \sH{32} \sV{11} $\\ 
   $+ \L{3 }  \ec{2}  \Nc{3} \bi{56} \sH{30}  \V{14}  \H{35} $  
 & $+ \L{3 }  \ec{2}  \p{12} \sH{15} \sH{16} \sV{11} \sV{22} $  
 & $+ \L{3 }  \ec{2}  \p{12} \sH{15} \sH{16}  \V{14}  \V{23} $\\ 
   $+ \L{3 }  \ec{2}  \p{12} \sH{16} \sH{38}  \V{15}  \V{35} $  
 & $+ \L{3 }  \ec{2}  \p{12} \sH{18} \sH{37}  \V{19}  \V{40} $  
 & $+ \L{3 }  \ec{2}  \p{13} \sV{12} \sV{21} \sV{31} \sV{32} $\\ 
   $+ \L{3 }  \ec{2}  \p{13} \sV{31} \sV{32}  \V{24}  \V{13} $  
 & $+ \L{3 }  \ec{2}  \p{23} \sH{15} \sH{22} \sH{30} \sH{37} $  
 & $+ \L{3 }  \ec{2}  \p{23} \sH{15} \sH{30}  \V{24}  \V{3 } $\\ 
   $+ \L{3 }  \ec{2}  \p{23} \sH{15} \sH{30}  \V{7 }  \V{25} $  
 & $+ \L{3 }  \ec{2}  \p{23} \sH{29} \sH{30} \sV{12} \sV{21} $  
 & $+ \L{3 }  \ec{2}  \p{23} \sH{29} \sH{30}  \V{24}  \V{13} $\\ 
   $+ \L{3 }  \ec{2} \bp{56} \sH{16} \sH{21} \sH{32} \sH{39} $  
 & $+ \L{3 }  \ec{2} \bp{56} \sH{16} \sH{21}  \H{42}  \H{35} $  
 & $+ \L{3 }  \ec{2} \bi{56} \sH{30} \sH{39} \sV{2 } \sV{31} $\\ 
   $+ \dots]$&& 
\end{tabular}\nolabel}
\eeqn
\newpage


\no Possible down--quark mass terms from $\vev{\h{4}\equiv\H{41}}$:
\beqn 
{\begin{tabular}{lll}
   $\pps\h{4}\, [ \Q{1 }  \dc{3} \sH{21} $  
 & $+ \Q{3 }  \dc{1} \sH{19} $&\\ 
   $+ \Q{2 }  \dc{2}  \p{23} \sH{38} $&&\\ 
   $+ \Q{1 }  \dc{3} \sH{15} \sH{18} \sH{19} $  
 & $+ \Q{2 }  \dc{2} \sH{38} \sV{31} \sV{32} $\\ 
   $+ \Q{1 }  \dc{1}  \Nc{1}  \p{12}  \H{26}  \V{7 } $  
 & $+ \Q{1 }  \dc{1}  \Nc{2}  \p{12}  \H{28}  \V{20} $  
 & $+ \Q{1 }  \dc{1}  \Nc{2}  \p{12}  \H{26}  \V{17} $\\ 
   $+ \Q{1 }  \dc{1}  \Nc{3}  \p{13} \sH{18} \sV{31} $  
 & $+ \Q{1 }  \dc{1}  \p{12} \sH{19} \sV{2 } \sV{21} $  
 & $+ \Q{1 }  \dc{1}  \p{12} \sH{19}  \V{9 }  \V{29} $\\ 
   $+ \Q{1 }  \dc{1}  \p{12} \sH{21} \sV{1 } \sV{22} $  
 & $+ \Q{1 }  \dc{1}  \p{12} \sH{38} \sV{1 } \sV{2 } $  
 & $+ \Q{1 }  \dc{1}  \p{12} \sH{38}  \V{4 }  \V{3 } $\\ 
   $+ \Q{1 }  \dc{1}  \p{12} \sH{38}  \V{5 }  \V{7 } $  
 & $+ \Q{1 }  \dc{2}  \Nc{1}  \p{12}  \H{28}  \V{20} $  
 & $+ \Q{1 }  \dc{2}  \Nc{1}  \p{12}  \H{26}  \V{17} $\\ 
   $+ \Q{1 }  \dc{2}  \p{12} \sH{21} \sV{11} \sV{22} $  
 & $+ \Q{1 }  \dc{2}  \p{12} \sH{21}  \V{14}  \V{23} $  
 & $+ \Q{1 }  \dc{2}  \p{12} \sH{38} \sV{2 } \sV{11} $\\
\end{tabular}\nolabel}
\eeqn
\newpage

\no Possible down--quark mass terms from $\vev{\h{4}\equiv\H{41}}$ continued:
\beqn 
{\begin{tabular}{lll}
   $+ \Q{3 }  \dc{2} \bp{13} \bpp{4} \sH{17} \sH{19} \sH{30} $  
 & $+ \Q{3 }  \dc{2} \sH{18} \sH{19} \sH{19}  \V{15}  \V{37} $  
 & $+ \Q{3 }  \dc{2} \sH{19}  \V{9 }  \V{40}  \V{17}  \V{37} $\\ 
   $+ \Q{3 }  \dc{2} \sH{21} \sH{30} \sH{39} \sV{22} \sV{31} $  
 & $+ \Q{3 }  \dc{2} \sH{21} \sH{30} \sV{31}  \H{42}  \V{23} $  
 & $+ \Q{3 }  \dc{2} \sH{30} \sH{38} \sH{39} \sV{2 } \sV{31} $\\ 
   $+ \Q{3 }  \dc{3}  \Nc{1} \sH{16} \sH{19}  \H{26}  \V{37} $  
 & $+ \Q{3 }  \dc{3}  \Nc{1} \sH{19} \sH{30} \sH{32} \sV{21} $  
 & $+ \Q{3 }  \dc{3}  \Nc{1} \sH{19} \sH{30}  \V{24}  \H{35} $\\ 
   $+ \Q{3 }  \dc{3}  \Nc{1} \sH{30} \sH{32} \sH{38} \sV{1 } $  
 & $+ \Q{3 }  \dc{3}  \Nc{2} \sH{21} \sH{30} \sH{36} \sV{31} $  
 & $+ \Q{3 }  \dc{3}  \Nc{2} \sH{30} \sH{32} \sH{38} \sV{11} $\\ 
   $+ \Q{3 }  \dc{3}  \Nc{2} \sH{30} \sH{38}  \V{14}  \H{35} $  
 & $+ \Q{3 }  \dc{3}  \Nc{3} \sH{16} \sH{38}  \H{26}  \V{37} $  
 & $+ \Q{3 }  \dc{3}  \Nc{3} \sH{30} \sH{32} \sH{38} \sV{21} $\\ 
   $+ \Q{3 }  \dc{3}  \Nc{3} \sH{30} \sH{38}  \V{24}  \H{35} $  
 & $+ \Q{3 }  \dc{3}  \p{4 } \bp{56} \sH{21} \sH{21} \sH{36} $  
 & $+ \Q{3 }  \dc{3} \bp{4 } \bi{56} \sH{19} \sH{19} \sH{36} $\\ 
   $+ \Q{3 }  \dc{3} \sH{15} \sH{16} \sH{19} \sH{20} \sH{38} $  
 & $+ \Q{3 }  \dc{3} \sH{15} \sH{19} \sH{30}  \V{20}  \V{29} $  
 & $+ \Q{3 }  \dc{3} \sH{15} \sH{19} \sH{30}  \V{17}  \V{25} $\\ 
   $+ \Q{3 }  \dc{3} \sH{15} \sH{30} \sH{38}  \V{14}  \V{3 } $  
 & $+ \Q{3 }  \dc{3} \sH{16} \sH{19} \sH{38}  \V{5 }  \V{37} $  
 & $+ \Q{3 }  \dc{3} \sH{17} \sH{21} \sH{30}  \V{19}  \V{30} $\\
\end{tabular}\nolabel}
\eeqn
\newpage

\no Possible electron mass terms from $\vev{\h{4}\equiv\H{41}}$:
\beqn 
{\begin{tabular}{lll}
   $\pps\h{4}\, [ \L{1 }  \ec{3} \bp{4 } \sH{19} $  
 & $+ \L{2 }  \ec{2}  \p{23} \sH{38} $  
 & $+ \L{3 }  \ec{1}  \p{4 } \sH{21} $\\ 
   $+ \L{1 }  \ec{3}  \P{2 } \bpp{4} \sH{19} $  
 & $+ \L{2 }  \ec{2} \sH{38} \sV{31} \sV{32} $\\ 
   $+ \L{1 }  \ec{1}  \Nc{1}  \p{12}  \H{26}  \V{7 } $  
 & $+ \L{1 }  \ec{1}  \Nc{2}  \p{12}  \H{28}  \V{20} $  
 & $+ \L{1 }  \ec{1}  \Nc{2}  \p{12}  \H{26}  \V{17} $\\ 
   $+ \L{1 }  \ec{1}  \Nc{3}  \p{13} \sH{18} \sV{31} $  
 & $+ \L{1 }  \ec{1}  \p{12} \sH{19} \sV{2 } \sV{21} $  
 & $+ \L{1 }  \ec{1}  \p{12} \sH{19}  \V{9 }  \V{29} $\\ 
   $+ \L{1 }  \ec{1}  \p{12} \sH{21} \sV{1 } \sV{22} $  
 & $+ \L{1 }  \ec{1}  \p{12} \sH{38} \sV{1 } \sV{2 } $  
 & $+ \L{1 }  \ec{1}  \p{12} \sH{38}  \V{4 }  \V{3 } $\\ 
   $+ \L{1 }  \ec{1}  \p{12} \sH{38}  \V{5 }  \V{7 } $  
 & $+ \L{1 }  \ec{2}  \p{12} \sH{21}  \V{19}  \V{30} $  
 & $+ \L{1 }  \ec{2}  \p{12} \sH{21}  \V{15}  \V{27} $\\ 
   $+ \L{1 }  \ec{2}  \p{12} \sH{38}  \V{4 }  \V{13} $  
 & $+ \L{1 }  \ec{2}  \p{12} \sH{38}  \V{9 }  \V{19} $  
 & $+ \L{2 }  \ec{1}  \Nc{3}  \p{56} \sH{30} \sV{31} $\\ 
   $+ \L{2 }  \ec{1}  \p{12} \sH{19}  \V{20}  \V{29} $  
 & $+ \L{2 }  \ec{1}  \p{12} \sH{19}  \V{17}  \V{25} $  
 & $+ \L{2 }  \ec{1}  \p{12} \sH{38}  \V{14}  \V{3 } $\\
\end{tabular}\nolabel}
\eeqn
\newpage

\no Possible electron mass terms from $\vev{\h{4}\equiv\H{41}}$ continued:
\beqn 
{\begin{tabular}{lll}
   $+ \L{3 }  \ec{1}  \Nc{2} \sH{17} \sH{18} \sH{30} \sV{31} $  
 & $+ \L{3 }  \ec{1}  \Nc{2} \sH{21} \sH{22} \sH{30} \sV{31} $  
 & $+ \L{3 }  \ec{1}  \Nc{3} \sH{16} \sH{21}  \H{28}  \V{40} $\\ 
   $+ \L{3 }  \ec{1}  \p{12} \bp{23} \sV{31}  \H{25}  \V{20} $  
 & $+ \L{3 }  \ec{1}  \p{12} \bp{23} \sV{31}  \H{23}  \V{17} $  
 & $+ \L{3 }  \ec{1}  \p{13}  \p{23} \sH{22} \sH{37} \sH{38} $\\ 
   $+ \L{3 }  \ec{1}  \p{13}  \p{23} \sH{38}  \V{24}  \V{3 } $  
 & $+ \L{3 }  \ec{1}  \p{13}  \p{23} \sH{38}  \V{7 }  \V{25} $  
 & $+ \L{3 }  \ec{1} \sH{17} \sH{21} \sH{30} \sV{1 } \sV{12} $\\ 
   $+ \L{3 }  \ec{1} \sH{17} \sH{21} \sH{30}  \V{7 }  \V{15} $  
 & $+ \L{3 }  \ec{1} \sH{17} \sH{30} \sV{31}  \H{25}  \V{9 } $  
 & $+ \L{3 }  \ec{1} \sH{19} \sH{21} \sH{31}  \V{19}  \V{40} $\\ 
   $+ \L{3 }  \ec{1} \sH{19} \sH{21}  \H{35}  \V{13}  \V{33} $  
 & $+ \L{3 }  \ec{1} \sH{19} \sH{31} \sV{31}  \H{23}  \V{37} $  
 & $+ \L{3 }  \ec{1} \sH{21} \sH{30} \sV{31}  \H{25}  \V{39} $\\ 
   $+ \L{3 }  \ec{1} \sH{29} \sH{30} \sV{31}  \H{25}  \V{20} $  
 & $+ \L{3 }  \ec{1} \sH{29} \sH{30} \sV{31}  \H{23}  \V{17} $  
 & $+ \L{3 }  \ec{2}  \p{13}  \p{23} \sH{38} \sV{12} \sV{21} $\\ 
   $+ \L{3 }  \ec{2}  \p{13}  \p{23} \sH{38}  \V{24}  \V{13} $  
 & $+ \L{3 }  \ec{2} \bp{13}  \p{4 } \sH{15} \sH{21} \sH{30} $  
 & $+ \L{3 }  \ec{2} \sH{15} \sH{30} \sV{31}  \H{25}  \V{20} $\\ 
   $+ \L{3 }  \ec{2} \sH{15} \sH{30} \sV{31}  \H{23}  \V{17} $  
 & $+ \L{3 }  \ec{2} \sH{16} \sH{19} \sH{21}  \V{15}  \V{37} $  
 & $+ \L{3 }  \ec{3}  \Nc{1} \sH{16} \sH{19}  \H{26}  \V{37} $\\
\end{tabular}\nolabel}
\eeqn
\newpage

\section{Potential FNY Higgs Doublet Mass Terms (through
8th order)} 

\no Possible Higgs $\h{1}\hb{1}$ mass terms (through $8^{\rm th}$ order):
\beqn 
{\begin{tabular}{llll}
   $\pps\h{1}\bh{1} [ 
      \Nc{3} \sH{30} \sH{32} \sV{21} $
 & $+ \Nc{3} \sH{30}  \V{24}  \H{35} $  
 & $+\sH{30} \sH{32} \sH{37} \sH{39} $  
 & $+\sH{30} \sH{37}  \H{42}  \H{35} $\\
   $+ \p{23} \sH{29} \sH{30} \sH{36} \sH{37} $  
 & $+ \p{23} \sH{29} \sH{30} \sV{21} \sV{22} $  
 & $+ \p{23} \sH{29} \sH{30}  \V{24}  \V{23} $  
 & $+ \p{23} \sH{29} \sH{30}  \V{29}  \V{30} $\\
   $+ \p{23} \sH{29} \sH{30}  \V{25}  \V{27} $  
 & $+\bp{56} \sH{31} \sH{32} \sH{38} \sH{39} $  
 & $+\bp{56} \sH{31} \sH{38}  \H{42}  \H{35} $\\ 
   $+ \Nc{3} \sH{15} \sH{30} \sH{31} \sH{36} \sV{31} $  
 & $+ \Nc{3} \sH{30} \sH{31} \sV{31}  \V{25}  \V{37} $  
 & $+ \p{23} \bp{56} \sH{29} \sH{31} \sH{36} \sH{38} $  
 & $+ \p{23} \bp{56}  \H{25}  \H{25}  \H{26}  \H{26} $\\ 
   $+ \p{23} \bp{56}  \H{25}  \H{28}  \H{23}  \H{26} $  
 & $+ \p{23} \bp{56}  \H{28}  \H{28}  \H{23}  \H{23} $  
 & $+\sH{15} \sH{30} \sH{31} \sH{39} \sV{22} \sV{31} $  
 & $+\sH{15} \sH{30} \sH{31} \sV{31}  \H{42}  \V{23} $\\ 
   $+\sH{15} \sH{30} \sH{39} \sV{12}  \H{26}  \V{7 } $  
 & $+\sH{15} \sH{30}  \H{42}  \V{13}  \H{26}  \V{7 } $  
 & $+\sH{29} \sH{30} \sH{36} \sH{37} \sV{31} \sV{32} $  
 & $+\sH{29} \sH{30} \sV{21} \sV{22} \sV{31} \sV{32} $\\ 
   $+\sH{29} \sH{30} \sV{31} \sV{32}  \V{24}  \V{23} $  
 & $+\sH{29} \sH{30} \sV{31} \sV{32}  \V{29}  \V{30} $  
 & $+\sH{29} \sH{30} \sV{31} \sV{32}  \V{25}  \V{27} $  
 & $+ \dots]$\\ 
\end{tabular}\nolabel}
\eeqn
\newpage

\no Possible Higgs $\h{1}\hb{2}$ mass terms (through $8^{\rm th}$ order):
\beqn 
{\begin{tabular}{llll}
   $\pps\h{1}\bh{2} [  \p{12}$
 & $+ \dots ]$ 
\end{tabular}\nolabel}
\eeqn
\newpage


\no Possible Higgs $\h{1}\hb{3}$ mass terms (through $8^{\rm th}$ order):
\beqn 
{\begin{tabular}{llll}
   $\pps\h{1}\bh{3} [  \p{13} $\\
   $+ \Nc{1} \sH{29} \sH{30} \sH{30} \sH{32} \sV{1 } $  
 & $+ \Nc{2} \sH{17} \sH{30} \sH{30}  \V{4 }  \H{35} $  
 & $+ \Nc{2} \sH{19} \sH{30} \sH{32}  \V{34}  \H{35} $  
 & $+ \Nc{3} \sH{15} \sH{30} \sH{31}  \H{28}  \V{40} $\\ 
   $+ \Nc{3} \sH{16} \sH{29} \sH{30}  \H{26}  \V{37} $  
 & $+ \Nc{3} \sH{29} \sH{30} \sH{30} \sH{32} \sV{21} $  
 & $+ \Nc{3} \sH{29} \sH{30} \sH{30}  \V{24}  \H{35} $  
 & $+\sH{15} \sH{29} \sH{30} \sH{30}  \V{14}  \V{3 } $\\ 
   $+\sH{17} \sH{29} \sH{30} \sH{30}  \V{4 }  \V{13} $  
 & $+\sH{17} \sH{29} \sH{30} \sH{30}  \V{9 }  \V{19} $  
 & $+\sH{19} \sH{29} \sH{30} \sH{31}  \V{15}  \V{37} $  
 & $+\sH{19} \sH{29} \sH{30} \sH{32}  \V{34}  \V{13} $\\ 
   $+\sH{19} \sH{29} \sH{30} \sV{12}  \V{34}  \H{35} $  
 & $+\sH{21} \sH{29} \sH{30} \sH{30}  \V{19}  \V{39} $  
 & $+\sH{29} \sH{30} \sH{30} \sH{32} \sH{37} \sH{39} $  
 & $+\sH{29} \sH{30} \sH{30} \sH{37}  \H{42}  \H{35} $\\ 
   $+\sH{29} \sH{30} \sH{30} \sV{31}  \H{23}  \V{35} $  
 & $+\sH{29} \sH{30} \sH{37} \sV{32}  \H{28}  \V{40} $  
 & $+ \dots ]$\\ 
\end{tabular}\nolabel}
\eeqn
\newpage

\no Possible Higgs $\h{1}\hb{4}$ mass terms (through $8^{\rm th}$ order):
\beqn 
{\begin{tabular}{llll}
   $\pps\h{1}\bh{4} [ \sH{29} \sH{30} \sH{31} \sV{31} \sV{32} $\\ 
   $+ \Nc{1}  \p{12} \sH{30} \sH{31} \sH{32} \sV{1 } $  
 & $+ \Nc{2}  \p{12} \sH{18} \sH{30} \sH{32} \sV{1 } $  
 & $+ \Nc{3}  \p{12} \sH{16} \sH{31}  \H{26}  \V{37} $  
 & $+ \Nc{3}  \p{12} \sH{30} \sH{31} \sH{32} \sV{21} $\\ 
   $+ \Nc{3}  \p{12} \sH{30} \sH{31}  \V{24}  \H{35} $  
 & $+ \p{12} \sH{15} \sH{30} \sH{31}  \V{14}  \V{3 } $  
 & $+ \p{12} \sH{16} \sH{29} \sH{30} \sV{2 } \sV{11} $  
 & $+ \p{12} \sH{16} \sH{29} \sH{30}  \V{5 }  \V{17} $\\ 
   $+ \p{12} \sH{16} \sH{29} \sV{32}  \H{28}  \V{30} $  
 & $+ \p{12} \sH{16} \sH{29} \sV{32}  \H{26}  \V{27} $  
 & $+ \p{12} \sH{17} \sH{30} \sH{31}  \V{4 }  \V{13} $  
 & $+ \p{12} \sH{17} \sH{30} \sH{31}  \V{9 }  \V{19} $\\ 
   $+ \p{12} \sH{18} \sH{29} \sH{30} \sV{1 } \sV{12} $  
 & $+ \p{12} \sH{18} \sH{29} \sH{30}  \V{7 }  \V{15} $  
 & $+ \p{12} \sH{19} \sH{31} \sH{31}  \V{15}  \V{37} $  
 & $+ \p{12} \sH{19} \sH{31} \sH{32}  \V{34}  \V{13} $\\ 
   $+ \p{12} \sH{19} \sH{31} \sV{12}  \V{34}  \H{35} $  
 & $+ \p{12} \sH{21} \sH{30} \sH{31}  \V{19}  \V{39} $  
 & $+ \p{12} \sH{30} \sH{31} \sH{32} \sH{37} \sH{39} $  
 & $+ \p{12} \sH{30} \sH{31} \sH{37}  \H{42}  \H{35} $\\ 
   $+ \p{12} \sH{30} \sH{31} \sV{31}  \H{23}  \V{35} $  
 & $+ \p{12} \sH{30} \sH{32} \sV{1 }  \H{25}  \V{9 } $  
 & $+ \p{12} \sH{30}  \V{4 }  \H{35}  \H{23}  \V{7 } $  
 & $+ \p{12} \sH{31} \sH{37} \sV{32}  \H{28}  \V{40} $\\ 
   $+ \dots ]$ 
\end{tabular}\nolabel}
\eeqn
\newpage
  
\no Possible Higgs $\h{2}\hb{1}$ mass terms (through $8^{\rm th}$ order):
\beqn 
{\begin{tabular}{llll}
   $\pps\h{2}\bh{1} [ \bp{12} $\\
   $+ \Nc{1} \sH{30} \sH{32} \sH{36} \sH{37} \sV{1 } $  
 & $+ \Nc{1} \sH{30} \sH{32} \sV{1 } \sV{21} \sV{22} $  
 & $+ \Nc{1} \sH{30} \sH{32} \sV{1 }  \V{24}  \V{23} $  
 & $+ \Nc{1} \sH{30} \sH{32} \sV{1 }  \V{29}  \V{30} $\\ 
   $+ \Nc{1} \sH{30} \sH{32} \sV{1 }  \V{25}  \V{27} $  
 & $+ \Nc{1} \sH{30} \sV{1 } \sV{22}  \V{24}  \H{35} $  
 & $+ \Nc{1} \sH{30} \sV{12} \sV{21}  \H{28}  \V{39} $  
 & $+ \Nc{1} \sH{30}  \V{24}  \V{13}  \H{28}  \V{39} $\\ 
   $+ \Nc{1} \sH{31}  \H{28}  \V{19}  \V{30}  \V{40} $  
 & $+ \Nc{1} \sH{31}  \H{28}  \V{40}  \V{15}  \V{27} $  
 & $+ \Nc{1} \sH{31}  \V{19}  \V{40}  \H{26}  \V{27} $  
 & $+ \Nc{1}  \H{35}  \V{13}  \V{33}  \H{28}  \V{30} $\\ 
   $+ \Nc{1}  \H{35}  \V{13}  \V{33}  \H{26}  \V{27} $  
 & $+ \Nc{2} \sH{16} \sV{11} \sV{22}  \H{26}  \V{37} $  
 & $+ \Nc{2} \sH{16}  \V{14}  \V{23}  \H{26}  \V{37} $  
 & $+ \Nc{2} \sH{18}  \H{28}  \V{19}  \V{30}  \V{40} $\\ 
   $+ \Nc{2} \sH{18}  \H{28}  \V{40}  \V{15}  \V{27} $  
 & $+ \Nc{2} \sH{18}  \V{19}  \V{40}  \H{26}  \V{27} $  
 & $+ \Nc{2} \sH{30} \sH{36} \sV{2 } \sV{21} \sV{31} $  
 & $+ \Nc{2} \sH{30} \sH{36} \sV{31}  \V{9 }  \V{29} $\\ 
   $+ \Nc{3} \sH{15} \sH{16} \sH{36}  \H{28}  \V{30} $  
 & $+ \Nc{3} \sH{15} \sH{16} \sH{36}  \H{26}  \V{27} $  
 & $+ \Nc{3} \sH{16} \sH{36} \sH{37}  \H{26}  \V{37} $  
 & $+ \Nc{3} \sH{16} \sV{21} \sV{22}  \H{26}  \V{37} $\\ 
   $+ \Nc{3} \sH{16}  \V{24}  \V{23}  \H{26}  \V{37} $  
 & $+ \Nc{3} \sH{16}  \H{28}  \V{30}  \V{25}  \V{37} $  
 & $+ \Nc{3} \sH{16}  \V{29}  \V{30}  \H{26}  \V{37} $  
 & $+ \Nc{3} \sH{16}  \H{26}  \V{25}  \V{27}  \V{37} $\\
\end{tabular}\nolabel}
\eeqn
\newpage

\no Possible Higgs $\h{2}\hb{1}$ mass terms continued:
\beqn 
{\begin{tabular}{llll}
   $+\sH{37} \sV{32}  \V{29}  \V{40}  \H{26}  \V{27} $  
 & $+\sH{38} \sV{1 } \sV{12} \sV{22}  \V{34}  \H{35} $  
 & $+\sH{38} \sV{12}  \V{34}  \V{13}  \H{28}  \V{39} $  
 & $+\sH{38} \sV{22}  \V{34}  \H{35}  \V{7 }  \V{15} $\\ 
   $+\sH{38}  \V{4 }  \H{35}  \V{13}  \V{23}  \V{33} $  
 & $+\sH{38}  \H{35}  \V{13}  \V{33}  \V{5 }  \V{27} $  
 & $+\sH{38}  \H{35}  \V{23}  \V{33}  \V{9 }  \V{19} $  
 & $+\sH{38}  \V{13}  \V{13}  \V{33}  \H{26}  \V{35} $\\ 
   $+\sV{1 } \sV{12}  \H{25}  \V{30}  \H{26}  \V{37} $  
 & $+\sV{1 } \sV{12}  \H{28}  \V{30}  \H{23}  \V{37} $  
 & $+\sV{1 } \sV{12}  \H{23}  \H{26}  \V{27}  \V{37} $  
 & $+\sV{22} \sV{31}  \V{34}  \H{35}  \H{25}  \V{30} $\\ 
   $+\sV{22} \sV{31}  \V{34}  \H{35}  \H{23}  \V{27} $  
 & $+ \V{4 }  \V{13}  \H{25}  \H{28}  \V{30}  \V{40} $  
 & $+ \V{4 }  \V{13}  \H{25}  \V{40}  \H{26}  \V{27} $  
 & $+ \V{4 }  \V{13}  \H{28}  \V{40}  \H{23}  \V{27} $\\ 
   $+ \H{25}  \H{28}  \V{9 }  \V{19}  \V{30}  \V{40} $  
 & $+ \H{25}  \H{28}  \V{9 }  \V{40}  \V{15}  \V{27} $  
 & $+ \H{25}  \V{9 }  \V{19}  \V{40}  \H{26}  \V{27} $  
 & $+ \H{25}  \V{30}  \H{26}  \V{7 }  \V{15}  \V{37} $\\ 
   $+ \H{28}  \V{9 }  \V{19}  \V{40}  \H{23}  \V{27} $  
 & $+ \H{28}  \V{30}  \H{23}  \V{7 }  \V{15}  \V{37} $  
 & $+ \V{19}  \V{30}  \H{23}  \H{26}  \V{7 }  \V{37} $  
 & $+ \H{23}  \H{26}  \V{7 }  \V{15}  \V{27}  \V{37} $\\ 
   $+ \dots ]$ 
\end{tabular}\nolabel}
\eeqn
\newpage

\no Possible Higgs $\h{2}\hb{2}$ mass terms (through $8^{\rm th}$ order):
\beqn 
{\begin{tabular}{llll}
   $\pps\h{2}\bh{2} [ \Nc{3} \sH{16}  \H{28}  \V{37} $  
 & $+\sH{37} \sV{32}  \H{28}  \V{40} $\\ 
   $+ \Nc{3} \bp{4 } \sH{16}  \H{28}  \V{40} $  
 & $+ \p{13} \sH{36} \sH{37} \sV{31} \sV{32} $  
 & $+ \p{13} \sV{21} \sV{22} \sV{31} \sV{32} $  
 & $+ \p{13} \sV{31} \sV{32}  \V{24}  \V{23} $\\
   $+ \p{13} \sV{31} \sV{32}  \V{29}  \V{30} $  
 & $+ \p{13} \sV{31} \sV{32}  \V{25}  \V{27} $  
 & $+ \p{4 } \sH{15} \sV{32}  \H{28}  \V{30} $  
 & $+ \p{4 } \sH{15} \sV{32}  \H{26}  \V{27} $\\
   $+ \p{4 } \sH{37} \sV{32}  \H{26}  \V{37} $\\ 
   $+ \Nc{1} \sH{16} \sH{20} \sH{29} \sH{30} \sV{31} $  
 & $+ \Nc{1} \sH{30} \sH{32} \sV{31}  \V{34}  \V{3 } $  
 & $+ \Nc{3}  \p{13} \bp{4 } \sH{16} \sH{36} \sV{31} $  
 & $+ \Nc{3} \sH{15} \sH{30} \sH{31} \sH{36} \sV{31} $\\ 
   $+ \Nc{3} \sH{30} \sH{31} \sV{31}  \V{25}  \V{37} $  
 & $+ \p{13} \bp{4 } \sH{16} \sH{39} \sV{22} \sV{31} $  
 & $+ \p{13} \bp{4 } \sH{16} \sV{31}  \H{42}  \V{23} $  
 & $+\sH{15} \sH{18} \sH{19} \sH{30}  \V{19}  \V{39} $\\ 
   $+\sH{15} \sH{19} \sH{22} \sH{30}  \V{4 }  \V{13} $  
 & $+\sH{15} \sH{19} \sH{22} \sH{30}  \V{9 }  \V{19} $  
 & $+\sH{15} \sH{20} \sH{30} \sV{31}  \H{23}  \V{7 } $  
 & $+\sH{15} \sH{30} \sH{31} \sH{39} \sV{22} \sV{31} $\\ 
   $+\sH{15} \sH{30} \sH{31} \sV{31}  \H{42}  \V{23} $  
 & $+\sH{15} \sH{30} \sH{39} \sV{12}  \H{26}  \V{7 } $  
 & $+\sH{15} \sH{30} \sV{1 } \sV{32}  \V{14}  \V{33} $  
 & $+\sH{15} \sH{30}  \H{42}  \V{13}  \H{26}  \V{7 } $\\
\end{tabular}\nolabel}
\eeqn
\newpage


\no Possible Higgs $\h{2}\hb{3}$ mass terms (through $8^{\rm th}$ order):
\beqn 
{\begin{tabular}{llll}
   $\h{2}\bh{3} [ \bp{23} $&&&\\
   $+ \Nc{1} \sH{30} \sH{32} \sV{1 }  \V{34}  \V{33} $  
 & $+ \Nc{2} \sH{15} \sH{22} \sH{30}  \H{28}  \V{30} $  
 & $+ \Nc{2} \sH{15} \sH{22} \sH{30}  \H{26}  \V{27} $  
 & $+ \Nc{2} \sH{15} \sH{30} \sH{36}  \H{26}  \V{7 } $\\ 
   $+ \Nc{2} \sH{22} \sH{30} \sH{37}  \H{26}  \V{37} $  
 & $+ \Nc{2} \sH{30} \sV{2 } \sV{21}  \H{28}  \V{40} $  
 & $+ \Nc{2} \sH{30}  \V{24}  \V{3 }  \H{26}  \V{37} $  
 & $+ \Nc{2} \sH{30}  \H{28}  \V{9 }  \V{29}  \V{40} $\\ 
   $+ \Nc{2} \sH{30}  \H{26}  \V{7 }  \V{25}  \V{37} $  
 & $+ \Nc{3} \sH{16}  \V{34}  \V{33}  \H{26}  \V{37} $  
 & $+ \Nc{3} \sH{30} \sH{32} \sV{21}  \V{34}  \V{33} $  
 & $+ \Nc{3} \sH{30}  \V{24}  \V{34}  \H{35}  \V{33} $\\ 
   $+\sH{15} \sH{30}  \V{14}  \V{34}  \V{3 }  \V{33} $  
 & $+\sH{15} \sH{30}  \H{25}  \H{28}  \V{30}  \V{39} $  
 & $+\sH{15} \sH{30}  \H{25}  \V{39}  \H{26}  \V{27} $  
 & $+\sH{15} \sH{30}  \H{28}  \V{39}  \H{23}  \V{27} $\\ 
   $+\sH{17} \sH{30} \sH{30} \sH{37}  \V{25}  \V{35} $  
 & $+\sH{17} \sH{30} \sH{32} \sH{37}  \V{34}  \V{23} $  
 & $+\sH{17} \sH{30} \sH{37} \sV{22}  \V{34}  \H{35} $  
 & $+\sH{17} \sH{30}  \V{4 }  \V{34}  \V{13}  \V{33} $\\ 
   $+\sH{17} \sH{30}  \V{34}  \V{33}  \V{9 }  \V{19} $  
 & $+\sH{19} \sH{31}  \V{34}  \V{33}  \V{15}  \V{37} $  
 & $+\sH{19} \sH{32}  \V{34}  \V{34}  \V{13}  \V{33} $  
 & $+\sH{19} \sV{12}  \V{34}  \V{34}  \H{35}  \V{33} $\\ 
   $+\sH{21} \sH{30}  \V{34}  \V{33}  \V{19}  \V{39} $  
 & $+\sH{29} \sH{31}  \V{30}  \V{40}  \V{27}  \V{37} $  
 & $+\sH{30} \sH{32} \sH{37} \sH{39}  \V{34}  \V{33} $  
 & $+\sH{30} \sH{37}  \H{42}  \V{34}  \H{35}  \V{33} $\\
\end{tabular}\nolabel}
\eeqn
\newpage

\no Possible Higgs $\h{2}\hb{4}$ mass terms (through $8^{\rm th}$ order):
\beqn 
{\begin{tabular}{llll}
   $\pps\h{2}\bh{4} [\sH{31} $\\
   $+\sH{16} \sH{19} \sV{2 }  \V{34}  \H{35} $  
 & $+\sH{16} \sH{29} \sV{32}  \H{28}  \V{30} $  
 & $+\sH{16} \sH{29} \sV{32}  \H{26}  \V{27} $\\ 
   $+ \Nc{2}  \p{12} \sH{16} \sH{36}  \H{28}  \V{9 } $  
 & $+ \Nc{2}  \p{13} \sH{16} \sH{18} \sH{20} \sV{31} $  
 & $+ \Nc{3} \bi{56} \sH{18} \sH{30}  \H{28}  \V{39} $  
 & $+ \Nc{3} \bi{56} \sH{22} \sH{30}  \H{28}  \V{9 } $\\ 
   $+ \p{12} \sH{16} \sV{2 } \sV{11}  \V{34}  \V{33} $  
 & $+ \p{12} \sH{16}  \V{34}  \V{33}  \V{5 }  \V{17} $  
 & $+ \p{12} \sH{16}  \H{25}  \V{30}  \H{26}  \V{35} $  
 & $+ \p{12} \sH{16}  \H{28}  \V{30}  \H{23}  \V{35} $\\ 
   $+ \p{12} \sH{16}  \H{23}  \H{26}  \V{27}  \V{35} $  
 & $+ \p{12} \sH{18} \sH{30} \sH{37}  \V{29}  \V{39} $  
 & $+ \p{12} \sH{18} \sH{37}  \H{35}  \V{23}  \V{33} $  
 & $+ \p{12} \sH{18} \sV{1 } \sV{12}  \V{34}  \V{33} $\\ 
   $+ \p{12} \sH{18}  \V{34}  \V{33}  \V{7 }  \V{15} $  
 & $+ \p{12} \sH{22} \sH{30} \sH{37} \sV{2 } \sV{21} $  
 & $+ \p{12} \sH{22} \sH{30} \sH{37}  \V{9 }  \V{29} $  
 & $+ \p{12} \sH{30} \sV{2 } \sV{21}  \V{24}  \V{3 } $\\ 
   $+ \p{12} \sH{30} \sV{2 } \sV{21}  \V{7 }  \V{25} $  
 & $+ \p{12} \sH{30}  \V{24}  \V{3 }  \V{9 }  \V{29} $  
 & $+ \p{12} \sH{30}  \V{9 }  \V{29}  \V{7 }  \V{25} $  
 & $+ \p{13} \sH{16} \sH{18} \sH{19}  \V{15}  \V{35} $\\ 
   $+ \p{13} \sH{16} \sH{20} \sH{21} \sV{1 } \sV{12} $  
 & $+ \p{13} \sH{16} \sH{20} \sH{21}  \V{7 }  \V{15} $  
 & $+ \p{13} \sH{16} \sH{20} \sV{31}  \H{25}  \V{9 } $  
 & $+ \p{13} \sH{16} \sH{39} \sV{12}  \H{28}  \V{9 } $\\
\end{tabular}\nolabel}
\eeqn
\newpage

\no Possible Higgs $\h{3}\hb{1}$ mass terms (through $8^{\rm th}$ order):
\beqn 
{\begin{tabular}{llll}
 $\h{3}\bh{1} [ \bp{13} $\\ 
   $+ \Nc{1} \sH{30} \sH{32} \sV{1 } \sV{11} \sV{12} $  
 & $+ \Nc{1} \sH{30} \sH{32} \sV{1 }  \V{14}  \V{13} $  
 & $+ \Nc{1} \sH{30} \sH{32} \sV{1 }  \V{19}  \V{20} $  
 & $+ \Nc{1} \sH{30} \sH{32} \sV{1 }  \V{15}  \V{17} $\\ 
   $+ \Nc{1} \sH{30} \sH{32} \sV{11}  \V{7 }  \V{15} $  
 & $+ \Nc{1} \sH{30} \sV{1 } \sV{12}  \V{14}  \H{35} $  
 & $+ \Nc{1} \sH{30}  \V{14}  \H{35}  \V{7 }  \V{15} $  
 & $+ \Nc{3} \sH{16} \sV{11} \sV{12}  \H{26}  \V{37} $\\ 
   $+ \Nc{3} \sH{16}  \V{14}  \V{13}  \H{26}  \V{37} $  
 & $+ \Nc{3} \sH{16}  \H{28}  \V{20}  \V{15}  \V{37} $  
 & $+ \Nc{3} \sH{16}  \V{19}  \V{20}  \H{26}  \V{37} $  
 & $+ \Nc{3} \sH{16}  \H{26}  \V{15}  \V{17}  \V{37} $\\ 
   $+ \Nc{3} \sH{30} \sH{32} \sV{11} \sV{12} \sV{21} $  
 & $+ \Nc{3} \sH{30} \sH{32} \sV{11}  \V{24}  \V{13} $  
 & $+ \Nc{3} \sH{30} \sH{32} \sV{21}  \V{14}  \V{13} $  
 & $+ \Nc{3} \sH{30} \sH{32} \sV{21}  \V{19}  \V{20} $\\ 
   $+ \Nc{3} \sH{30} \sH{32} \sV{21}  \V{15}  \V{17} $  
 & $+ \Nc{3} \sH{30} \sV{11} \sV{12}  \V{24}  \H{35} $  
 & $+ \Nc{3} \sH{30} \sV{12} \sV{21}  \V{14}  \H{35} $  
 & $+ \Nc{3} \sH{30}  \V{14}  \V{24}  \H{35}  \V{13} $\\ 
   $+ \Nc{3} \sH{30}  \V{24}  \H{35}  \V{19}  \V{20} $  
 & $+ \Nc{3} \sH{30}  \V{24}  \H{35}  \V{15}  \V{17} $  
 & $+\sH{15} \sH{30} \sV{11} \sV{12}  \V{14}  \V{3 } $  
 & $+\sH{15} \sH{30}  \V{14}  \V{14}  \V{3 }  \V{13} $\\ 
   $+\sH{15} \sH{30}  \V{14}  \V{3 }  \V{19}  \V{20} $  
 & $+\sH{15} \sH{30}  \V{14}  \V{3 }  \V{15}  \V{17} $  
 & $+\sH{17} \sH{30} \sV{11} \sV{12}  \V{4 }  \V{13} $  
 & $+\sH{17} \sH{30} \sV{11} \sV{12}  \V{9 }  \V{19} $\\
\end{tabular}\nolabel}
\eeqn
\newpage

\no Possible Higgs $\h{3}\hb{2}$ mass terms (through $8^{\rm th}$ order):
\beqn 
{\begin{tabular}{llll}
   $\pps\h{3}\bh{2} [  \p{23} $\\
   $+ \Nc{1} \sH{30} \sH{32} \sV{1 } \sV{31} \sV{32} $  
 & $+ \Nc{3} \sH{16} \sV{31} \sV{32}  \H{26}  \V{37} $  
 & $+ \Nc{3} \sH{30} \sH{32} \sV{21} \sV{31} \sV{32} $  
 & $+ \Nc{3} \sH{30} \sV{31} \sV{32}  \V{24}  \H{35} $\\ 
   $+\sH{15} \sH{30} \sV{31} \sV{32}  \V{14}  \V{3 } $  
 & $+\sH{17} \sH{30} \sV{31} \sV{32}  \V{4 }  \V{13} $  
 & $+\sH{17} \sH{30} \sV{31} \sV{32}  \V{9 }  \V{19} $  
 & $+\sH{19} \sH{31} \sV{31} \sV{32}  \V{15}  \V{37} $\\ 
   $+\sH{19} \sH{32} \sV{31} \sV{32}  \V{34}  \V{13} $  
 & $+\sH{19} \sV{12} \sV{31} \sV{32}  \V{34}  \H{35} $  
 & $+\sH{21} \sH{30} \sV{31} \sV{32}  \V{19}  \V{39} $  
 & $+\sH{30} \sH{32} \sH{37} \sH{39} \sV{31} \sV{32} $\\ 
   $+\sH{30} \sH{37} \sV{31} \sV{32}  \H{42}  \H{35} $  
 & $+\sH{30} \sV{31} \sV{31} \sV{32}  \H{23}  \V{35} $  
 & $+\sH{37} \sV{31} \sV{32} \sV{32}  \H{28}  \V{40} $  
 & $+ \dots ]$\\
\end{tabular}\nolabel}
\eeqn
\newpage

\no Possible Higgs $\h{3}\hb{3}$ mass terms (through $8^{\rm th}$ order):
\beqn 
{\begin{tabular}{llll}
   $\pps\h{3}\bh{3} [     \Nc{1} \sH{30} \sH{32} \sV{1 } $  
 & $+\sH{15} \sH{30}  \V{14}  \V{3 } $  
 & $+\sH{17} \sH{30}  \V{4 }  \V{13} $  
 & $+\sH{17} \sH{30}  \V{9 }  \V{19} $\\
   $+\sH{19} \sH{31}  \V{15}  \V{37} $  
 & $+\sH{19} \sH{32}  \V{34}  \V{13} $  
 & $+\sH{19} \sV{12}  \V{34}  \H{35} $  
 & $+\sH{21} \sH{30}  \V{19}  \V{39} $\\
   $+\sH{30} \sV{31}  \H{23}  \V{35} $\\ 
   $+ \Nc{1}  \pp{4} \sH{30}  \V{4 }  \H{35} $  
 & $+ \Nc{2} \bpp{4} \sH{22} \sH{30} \sV{31} $  
 & $+ \p{12} \sV{11} \sV{12}  \V{34}  \V{33} $  
 & $+ \p{12}  \V{14}  \V{34}  \V{13}  \V{33} $\\
   $+ \p{12}  \V{34}  \V{33}  \V{19}  \V{20} $  
 & $+ \p{12}  \V{34}  \V{33}  \V{15}  \V{17} $  
 & $+\bp{12} \sH{29} \sH{30} \sV{31} \sV{32} $  
 & $+ \pp{4} \sH{15} \sH{30} \sV{2 } \sV{11} $\\
   $+ \pp{4} \sH{15} \sH{30}  \V{5 }  \V{17} $  
 & $+ \pp{4} \sH{21} \sH{30}  \V{15}  \V{35} $  
 & $+\bpp{4} \sH{17} \sH{30} \sV{1 } \sV{12} $  
 & $+\bpp{4} \sH{17} \sH{30}  \V{7 }  \V{15} $\\
   $+\bpp{4} \sH{19} \sH{31}  \V{19}  \V{40} $  
 & $+\bpp{4} \sH{19}  \H{35}  \V{13}  \V{33} $  
 & $+\bpp{4} \sH{30} \sV{31}  \H{25}  \V{39} $\\ 
   $+ \Nc{1} \sH{16} \sH{20} \sH{29} \sH{30} \sV{31} $  
 & $+ \Nc{1} \sH{30} \sH{32} \sV{31}  \V{34}  \V{3 } $  
 & $+ \Nc{3} \sH{15} \sH{30} \sH{31} \sH{36} \sV{31} $  
 & $+ \Nc{3} \sH{30} \sH{31} \sV{31}  \V{25}  \V{37} $\\
\end{tabular}\nolabel}
\eeqn
\newpage

\no Possible Higgs $\h{3}\hb{4}$ mass terms (through $8^{\rm th}$ order):
\beqn 
{\begin{tabular}{llll}
   $\pps\h{3}\bh{3} [    \sH{31} \sV{31} \sV{32} $\\ 
   $+ \p{12} \sH{16} \sV{2 } \sV{11} $  
 & $+ \p{12} \sH{16}  \V{5 }  \V{17} $  
 & $+ \p{12} \sH{18} \sV{1 } \sV{12} $  
 & $+ \p{12} \sH{18}  \V{7 }  \V{15} $\\
   $+ \p{12}  \pp{4} \sH{18}  \V{4 }  \V{13} $  
 & $+ \p{12}  \pp{4} \sH{18}  \V{9 }  \V{19} $  
 & $+ \p{12} \bpp{4} \sH{16}  \V{14}  \V{3 } $  
 & $+\sH{16} \sH{19} \sV{32}  \V{4 }  \H{35} $\\
   $+ \Nc{1}  \p{12} \sH{16} \sH{20} \sH{31} \sV{31} $  
 & $+ \Nc{1}  \p{23} \sH{30} \sH{31} \sH{32} \sV{1 } $  
 & $+ \Nc{2}  \p{12} \sH{16} \sH{18} \sH{20} \sV{31} $  
 & $+ \Nc{2}  \p{23} \sH{18} \sH{30} \sH{32} \sV{1 } $\\ 
   $+ \Nc{3}  \p{23} \sH{16} \sH{31}  \H{26}  \V{37} $  
 & $+ \Nc{3}  \p{23} \sH{30} \sH{31} \sH{32} \sV{21} $  
 & $+ \Nc{3}  \p{23} \sH{30} \sH{31}  \V{24}  \H{35} $  
 & $+ \p{12} \sH{16} \sH{18} \sH{19}  \V{15}  \V{35} $\\ 
   $+ \p{12} \sH{16} \sH{20} \sH{21} \sV{1 } \sV{12} $  
 & $+ \p{12} \sH{16} \sH{20} \sH{21}  \V{7 }  \V{15} $  
 & $+ \p{12} \sH{16} \sH{20} \sV{31}  \H{25}  \V{9 } $  
 & $+ \p{12} \sH{16} \sH{39} \sV{12}  \H{28}  \V{9 } $\\ 
   $+ \p{12} \sH{16} \sV{11} \sV{32}  \V{4 }  \V{33} $  
 & $+ \p{12} \sH{16}  \H{42}  \V{13}  \H{28}  \V{9 } $  
 & $+ \p{12} \sH{16}  \V{9 }  \V{39}  \V{17}  \V{37} $  
 & $+ \p{12} \sH{16}  \V{9 }  \V{40}  \V{17}  \V{35} $\\ 
   $+ \p{12} \sH{18} \sV{12} \sV{31}  \V{34}  \V{3 } $  
 & $+ \p{12} \sV{12} \sV{21} \sV{32}  \H{26}  \V{7 } $  
 & $+ \p{12} \sV{32}  \V{24}  \V{13}  \H{26}  \V{7 } $  
 & $+ \p{13} \sH{16} \sV{2 } \sV{11} \sV{31} \sV{32} $\\
\end{tabular}\nolabel}
\eeqn
\newpage

\no Possible Higgs $\h{4}\hb{1}$ mass terms (through $8^{\rm th}$ order):
\beqn 
{\begin{tabular}{llll}
   $\pps\h{4}\bh{1} [    
     \sH{38} \sV{11} \sV{12} $  
 & $+\sH{38}  \V{14}  \V{13} $  
 & $+\sH{38}  \V{19}  \V{20} $  
 & $+\sH{38}  \V{15}  \V{17} $\\
   $+ \Nc{1} \sH{30} \sV{31}  \V{20}  \V{29} $  
 & $+ \Nc{1} \sH{30} \sV{31}  \V{17}  \V{25} $  
 & $+\sH{15}  \V{20}  \V{40}  \V{17}  \V{27} $  
 & $+\sH{15}  \V{30}  \V{40}  \V{17}  \V{17} $\\
   $+\sH{36} \sH{37} \sH{38} \sV{31} \sV{32} $  
 & $+\sH{37}  \V{20}  \V{40}  \V{17}  \V{37} $  
 & $+\sH{38} \sV{21} \sV{22} \sV{31} \sV{32} $  
 & $+\sH{38} \sV{31} \sV{32}  \V{24}  \V{23} $\\
   $+\sH{38} \sV{31} \sV{32}  \V{29}  \V{30} $  
 & $+\sH{38} \sV{31} \sV{32}  \V{25}  \V{27} $\\ 
   $+ \Nc{1} \bp{13} \sH{19} \sH{30} \sH{32} \sV{21} $  
 & $+ \Nc{1} \bp{13} \sH{19} \sH{30}  \V{24}  \H{35} $  
 & $+ \Nc{1} \bp{13} \sH{30} \sH{32} \sH{38} \sV{1 } $  
 & $+ \Nc{1}  \p{23} \sH{30} \sV{11}  \V{24}  \V{33} $\\ 
   $+ \Nc{1}  \p{23} \sH{30} \sV{21}  \V{14}  \V{33} $  
 & $+ \Nc{1}  \p{56} \sH{30} \sV{31}  \V{19}  \V{30} $  
 & $+ \Nc{1}  \p{56} \sH{30} \sV{31}  \V{15}  \V{27} $  
 & $+ \Nc{2} \bp{13} \sH{21} \sH{30} \sH{36} \sV{31} $\\ 
   $+ \Nc{3} \bp{13} \sH{16} \sH{38}  \H{26}  \V{37} $  
 & $+ \Nc{3} \bp{13} \sH{30} \sH{32} \sH{38} \sV{21} $  
 & $+ \Nc{3} \bp{13} \sH{30} \sH{38}  \V{24}  \H{35} $  
 & $+ \Nc{3} \bp{4 } \sH{16} \sH{36} \sH{38} \sV{31} $\\ 
   $+\bp{12} \sH{21} \sH{30} \sH{39} \sV{12} \sV{31} $  
 & $+\bp{12} \sH{21} \sH{30} \sV{31}  \H{42}  \V{13} $  
 & $+\bp{13} \sH{15} \sH{19} \sH{30}  \V{20}  \V{29} $  
 & $+\bp{13} \sH{15} \sH{19} \sH{30}  \V{17}  \V{25} $\\
\end{tabular}%
\nolabel}
\eeqn
\newpage

\section{Higgs Doublet, Quark, and Charged Lepton Mass Terms 
(through 9th order) for Flat Direction FDNA(5+8)} 

\no FDNA(5+8) fields with VEVs:\hfill

$\p{23}\p{4} \bp{4} \pp{4} \bpp{4} \bpp{56} \sH{15} \sH{21}$$\sH{30}$$\H{31}$$\sH{37}$$\sH{38}$$\Nc{3}$$\H{26}$$\V{5}$$\V{35}$$\V{37}$\hfill
\vskip 0.2truecm
\vskip 0.2truecm

\no Additional VEVs allowed below FI scale: $\bp{12}\sim 10^{-4}$
\vskip 0.2truecm
\vskip 0.2truecm
               
\no \underline{Higgs Doublet Mass Terms} 
\beqn 
{\begin{tabular}{lll}
   $\pps\h{2} \hb{1} \pb{12}$
 & $+ \h{2} \hb{4} \Hs{31}$   
 & $+ \h{3} \hb{2} \p{23} $\\  
   $+ \h{4} \hb{3} \Hs{38}$
 & $+ \h{4} \hb{4} \p{23} \H{31} \H{38}$
 & $+ \h{1} \hb{3} \Nc{3} \pp{4} \H{15} \H{30} \H{31} \H{26} \V{37}$ 
\end{tabular}
\label{ch}}
\eeqn

\vskip 0.2truecm
\no \underline{$\hb{}\Q{}\uc{}$ Terms} 
\beqn 
{\begin{tabular}{lll}
   $\pps \hb{1} \Q{1} \uc{1} $
 & $+    \hb{4} \Q{3} \uc{3} \sH{30} $ 
 & $+    \hb{3} \Q{1} \uc{2} \sH{15} \sH{30} $\\   
   $+    \hb{3} \Q{2} \uc{1} \pp{4}  \sH{15} \sH{30} $  
 & $+    \hb{4} \Q{1} \uc{2} \p{23}  \sH{15} \sH{30} \H{31}$
 & $+    \hb{3} \Q{1} \uc{2} \Nc{3} \bpp{56} \sH{30} \H{26} \V{37} $\\
   $+    \hb{4} \Q{2} \uc{1} \p{23} \pp{4} \sH{15} \sH{30} \sH{31} $
 & $+    \hb{3} \Q{2} \uc{1} \Nc{3} \pp{4} \bpp{56} \sH{30} \H{26} \V{37} $ 
 & $+    \hb{2} \Q{1} \uc{2} \Nc{3} \p{23} \bpp{56} \sH{30} \H{26} \V{37} $
\end{tabular}%
\label{ce}}
\eeqn

\vskip 0.2truecm
\no Note: no proton decay terms exist for this flat direction through
at least 9th order. 
  
\vskip 0.2truecm
\no \underline{Additional Higgs Doublet mass terms when $\vev{\bp{23},\pp{56}}>0$} 
\beqn 
{\begin{tabular}{llll}
   $\h{2} \hb{3} \bp{23}$
&\phantom{$\h{2} \hb{3} \bp{23}$}
&\phantom{$\h{2} \hb{3} \bp{23}$}
\end{tabular}
\label{ach}}
\eeqn

\vskip 0.2truecm
\no \underline{Additional $\h{}\Q{}\dc{}$ mass terms when $\vev{\bp{23},\pp{56}}>0$} 
\beqn 
{\begin{tabular}{llll}
  $\h{4} \Q{1} \dc{2} \p{23} \pp{56} \sH{15} \sH{31} \sH{38}$ 
 &$+\h{1} \Q{3} \dc{3} \pp{4} \pp{56} \sH{15} \sH{15} \sH{30} \sH{31}$ 
 &\phantom{$\h{1} \Q{3} \dc{3} \pp{4} \pp{56}$} 
\end{tabular}
\label{acd}}
\eeqn

\vskip 0.2truecm
\no \underline{Additional $\h{}\L{}\ec{}$ mass terms when $\vev{\bp{23},\pp{56}}>0$} 
\beqn 
{\begin{tabular}{llll}
  $ \h{1} \L{3} \ec{3} \pp{4} \pp{56} \sH{15} \sH{15} \sH{30} \sH{31}$
 &$ +\h{4} \L{2} \ec{1} \p{23} \pp{4}  \pp{56} \sH{15} \sH{31} \sH{38}$
 &\phantom{$\h{1} \Q{3} \dc{3} \pp{4} \pp{56}$} 
\end{tabular}
\label{ace}}
\eeqn

\newpage

\newpage
\def\AEF{A.E. Faraggi}
\def\AP#1#2#3{{\it Ann.\ Phys.}\/ {\bf#1} (#2) #3}
\def\NPB#1#2#3{{\it Nucl.\ Phys.}\/ {\bf B#1} (#2) #3}
\def\NPBPS#1#2#3{{\it Nucl.\ Phys.}\/ {{\bf B} (Proc. Suppl.) {\bf #1}} (#2) 
 #3}
\def\PLB#1#2#3{{\it Phys.\ Lett.}\/ {\bf B#1} (#2) #3}
\def\PRD#1#2#3{{\it Phys.\ Rev.}\/ {\bf D#1} (#2) #3}
\def\PRL#1#2#3{{\it Phys.\ Rev.\ Lett.}\/ {\bf #1} (#2) #3}
\def\PRT#1#2#3{{\it Phys.\ Rep.}\/ {\bf#1} (#2) #3}
\def\PTP#1#2#3{{\it Prog.\ Theo.\ Phys.}\/ {\bf#1} (#2) #3}
\def\MODA#1#2#3{{\it Mod.\ Phys.\ Lett.}\/ {\bf A#1} (#2) #3}
\def\MPLA#1#2#3{{\it Mod.\ Phys.\ Lett.}\/ {\bf A#1} (#2) #3}
\def\IJMP#1#2#3{{\it Int.\ J.\ Mod.\ Phys.}\/ {\bf A#1} (#2) #3}
\def\IJMPA#1#2#3{{\it Int.\ J.\ Mod.\ Phys.}\/ {\bf A#1} (#2) #3}
\def\nuvc#1#2#3{{\it Nuovo Cimento}\/ {\bf #1A} (#2) #3}
\def\RPP#1#2#3{{\it Rept.\ Prog.\ Phys.}\/ {\bf #1} (#2) #3}
\def\etal{{\it et al\/}}
               

\def\bibiteml#1#2{ }
\bibliographystyle{unsrt}

\hfill\vfill\eject
\end{document}